\documentclass[aps,prb,twocolumn,superscriptaddress,floatfix]{revtex4-1}

\usepackage{graphicx}
\usepackage{subfigure}
\usepackage{amsmath}
\usepackage{lineno,hyperref}
\usepackage{tikz}
\usepackage{physics}

\renewcommand{\vec}[1]{\mathbf{#1}}
\newcommand{\be}{\hat b}
\newcommand{\bed}{\hat b^\dagger}

\usepackage{xcolor}
\usepackage[final]{changes}
\definechangesauthor[color=blue]{common}
\definechangesauthor[color=red, name={Paulo Santos}]{PVS}
\definechangesauthor[color=blue, name={Ronen Rapaport}]{RR}
\definechangesauthor[color=purple, name={Misha Lemeshko}]{ML}
\definechangesauthor[color=teal, name={Colin Hubert}]{CH}
\definechangesauthor[color=green, name={Areg Ghazaryan}]{AG}
\setauthormarkup{}

\newcommand{\rPVS}[2]{\replaced[id=PVS]{#1}{#2}}
\newcommand{\aPVS}[1]{\added[id=PVS]{#1}}

\begin{document}


\title{Attractive interactions, molecular complexes, and polarons in coupled dipolar exciton fluids}


\author{C. Hubert}
\affiliation{Paul-Drude-Institut f{\"u}r Festk{\"o}rperelektronik im Forschungsverbund Berlin eV., Hausvogteiplatz 5-7, 10117 Berlin, Germany}

\author{K. Cohen}
\affiliation{Racah Institute of Physics, The Hebrew University of Jerusalem, Jerusalem 9190401, Israel}


\author{A. Ghazaryan}
\affiliation{IST Austria (Institute of Science and Technology Austria), Am Campus 1, 3400 Klosterneuburg, Austria}

\author{M. Lemeshko}
\affiliation{IST Austria (Institute of Science and Technology Austria), Am Campus 1, 3400 Klosterneuburg, Austria}

\author{R. Rapaport}
\affiliation{Racah Institute of Physics, The Hebrew University of Jerusalem, Jerusalem 9190401, Israel}

\author{P. V. Santos}
\email[]{santos@pdi-berlin.de}
\affiliation{Paul-Drude-Institut f{\"u}r Festk{\"o}rperelektronik im Forschungsverbund Berlin eV., Hausvogteiplatz 5-7, 10117 Berlin, Germany}


\date{\today}

\begin{abstract}

Dipolar (or spatially indirect) excitons (IXs) in semiconductor double quantum well (DQW) subjected to an electric field are neutral species with a dipole moment oriented perpendicular to the DQW plane. Here, we theoretically study interactions between IXs in  stacked DQW bilayers, where the dipolar coupling can be  either attractive or repulsive depending on the relative positions of the particles. By using microscopic band structure calculations  to determine the electronic states  forming the excitons, we show that the attractive dipolar interaction between stacked IXs deforms their electronic  wave function, thereby increasing the inter-DQW interaction energy and making the IX even more electrically polarizable.  Many-particle interaction  effects are addressed by considering the coupling between a single IX in one of the DQWs to a cloud of IXs in the other DQW, which is modeled either as a closed-packed lattice or as a continuum IX fluid. 
We find that the lattice model yields  IX interlayer binding energies decreasing with increasing lattice density. This behavior is due to the dominating role of the intra-DQW dipolar repulsion, which prevents more than one exciton from entering the attractive region of the inter-DQW coupling. Finally, both models shows that the single IX distorts the distribution of IXs in the adjacent DQW, thus inducing the formation of an IX dipolar polaron \aPVS{di-polaron}. While the interlayer binding energy reduces with IX density for lattice \aPVS{di-}polarons, the continuous polaron model predicts a non-monotonous dependence on density in semi-quantitative agreement with a recent experimental study  [cf. Hubert {\it et al.}, Phys. Rev. {\bf X}9, 021026 (2019)].

%
%
\end{abstract}


\maketitle

\section{Introduction}
\label{Introduction}

The dipolar coupling is the dominating interaction mechanism between charge-neutral species. This interaction  is presently receiving considerable attention due to its special properties.  For two equal collinear dipoles with dipole moment $p \bf \hat z$ oriented along the $\bf \hat z$ direction in a medium with dielectric constant $\varepsilon\varepsilon_0$, the dipolar interaction energy can be expressed as 

\begin{equation}
U_{dd}(\textbf{r}) = \frac{ p^2} {4\pi\varepsilon\varepsilon_0} \frac{\left( 1-3 \cos^2\theta \right)}{r^3}.
\label{EqUdd}
\end{equation}

\noindent Here, $\theta$ is the angle  between the unit vector $ \bf \hat z$  and  the vector  \textbf{r}  connecting the two dipoles. 
The dipolar interaction has a long-range decay  proportional to  $r^{-3}$ and is a non-monotonic function of $\theta$. Furthermore, the interaction strength depends not only on the magnitude of {\bf r} but also on its orientation\deleted{ of the dipole moments}. In particular, the interaction between equal dipoles changes from repulsive to attractive for angles  $\theta<\arccos{[1/\sqrt{3}]}$. This behavior contrasts to the Coulomb interaction between two equally charged  particles, which is always repulsive and only depends on the distance between the particles.  

The interplay between dipolar attraction and repulsion gives rises to interesting phenomena in dipolar fluids.\cite{Lahaye_RPP72_126401_09} A typical example in classical systems is the  formation of patterns in ferrofluids.\cite{Kadau_N530_194_16} Recently, several investigations have addressed dipolar interaction in ensembles of  ultra-cold magnetic atoms and dipolar molecules. The anisotropic character of dipolar interactions affects the stability and determines the shape of clouds of atomic dipolar\cite{Lahaye_N448_672_07, Lahaye_RPP72_126401_09}
Due to the attractive dipolar component, the dispersion relation for density fluctuations  in these systems can exhibit a roton-maxon feature characterized by a  local minimum for non-zero wave vectors.\cite{Santos_PRL90_250403_03,Chomaz_NP14_442_18} Under appropriate conditions, instabilities around these local minima can induce the symmetric break leading to the formation of self-organized droplet crystals.\cite{Kadau_N530_194_16} Recently, these systems have been also shown to exhibit supersolidity, the combination of spatial ordering with superfluidity.\cite{Donner_report_2019,Boettcher_PRX9_11051_19,Tanzi_PRL122_130405_19,Chomaz_PRX6_41039_16}

\begin{figure}[!tbhp]
\begin{center}
\includegraphics[width=\columnwidth,angle=0, clip]{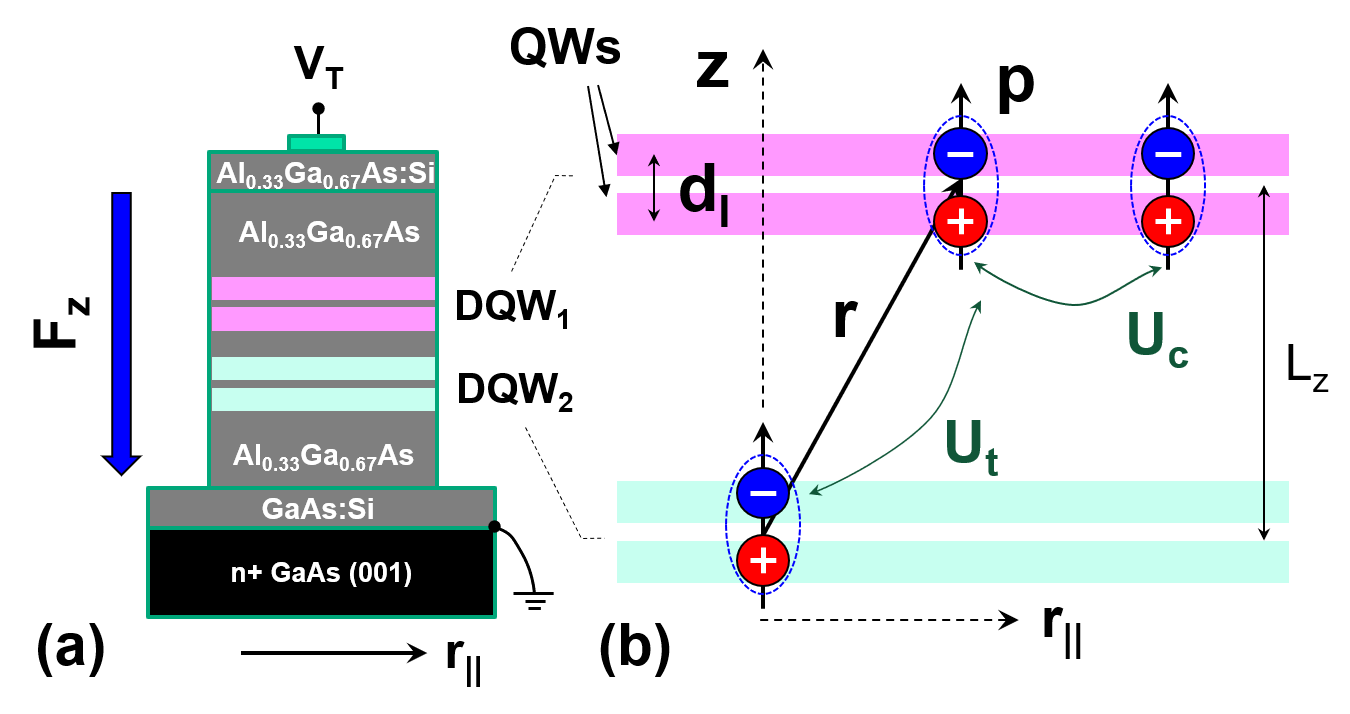} 
\end{center}
\caption{Dipolar interaction between indirect excitons (IXs) in stacked double quantum wells (DQWs). (a) DQWs consisting of GaAs QWs separated by a thin Al$_{0.3}$Ga$_{0.67}$As barrier are inserted within the intrinsic region of a reversed-biased Schottky diode subjected to a voltage $V_T$. The associated electric field $F_z$ induces the formation of IXs with dipole moment {\bf p} under optical excitation. (b) 
Intra-DQW ($U_c$) and inter-DQW ($U_t$) dipolar interaction  between two IXs. 
}
\label{DipInt}
\end{figure}

In the solid state, promising systems for the investigation of the coupling between mobile dipoles  include excitons in bi-layer electron systems~\cite{Eisenstein_N432_691_04,Eisenstein_ARCMP5_159_14} and spatially indirect excitons (IXs) in semiconductor double quantum well (DQW) structures (see Fig.~\ref{DipInt}a)~\cite{Butov_PRL92_117404_04,Rapaport_PRL92_117405_04,PVS177}. The latter consists of two QWs separated by a thin tunnel barrier (i.e., with a thickness much smaller than the  exciton Bohr radius). The DQWs are normally inserted within the intrinsic region of a reversed bias p-i-n junction to allow the application of a transverse electric field $F_\mathrm{z} \bf\hat{e}_z$, as shown for an (Al,Ga)As structure in the Fig.~\ref{DipInt}(a).  Field-induced tunneling drives photoexcited electrons and  holes to different QWs, thus creating excitons with a permanent dipole moment $p\bf\hat {e_z}$ oriented anti-parallel with respect to the field direction.

Most studies of IX dipoles have been restricted to particles confined  in a plane perpendicular to their dipole orientation. In this configuration,  the inter-IX interactions are only of repulsive nature. Recently, we have proposed and demonstrated a concept to access the attractive dipolar coupling between IXs using stacked DQWs [see Fig.~\ref{DipInt}(b)].\cite{PVS315} These configuration enables the orientation of dipoles with angles  $\theta<\arccos{[1/\sqrt{3}]}$. The barrier between the two DQW bilayers was chosen to be sufficiently large to prevent single-particle tunneling but small enough to allow for electrostatic coupling between them. Under these conditions, the attractive inter-DQW dipolar interaction  becomes sufficiently strong to induce density correlations and dragging between IX clouds in the two DQW bilayers. In addition, the inter-bilayer binding energies was determined from the energetic shifts of the photo-luminescence lines from the interacting IX clouds. Interestingly, the latter was found to increase with bilayer density up to densities of approx. $5\times 10^{10}$~cm$^{-2}$, where it reaches values exceeding 5 meV, and then to decrease for higher densities (cf. Fig.~6 of Ref.~\onlinecite{PVS315}). It is interesting to note that depending on density, the inter-layer binding can thus be  comparable to the intra-layer IX binding energy in GaAs QWs.

In a previous theoretical study,\cite{PVS284} we have calculated the inter-layer binding energy of an IX molecule consisting of two IXs, each in one of the DQW bilayer. This binding energy for the sample used in Ref.~\onlinecite{PVS315} reaches only a few tens of meV and is thus much smaller than the measured interlayer binding energies. The calculations were carried out using the point charge approximation (PCA), which assumes that the electrons and holes constituting the IX to be described as point charges located at the center of the QWs. They neglect, therefore, the internal charge distribution within the dipolar species. The PCA is a very good approximation for diluted gases of atomic and molecular dipoles, where the dipolar field can be considered slowly varying over the length of the dipoles.  It may fail, however, when the inter-dipolar distances become comparable to the length of the dipoles.
Also, we have previously not addressed  many-body effects arising from  many interacting dipoles in such a bilayer system

In this work, we first analyze the dipolar coupling beyond the PCA  approximation by addressing the coupling between extended dipoles with inter-dipolar separations comparable to the dipole lengths. For that purpose, we have  carried out self-consistent calculations of the energy and wave functions of electrons and holes for the system of two interacting IXs shown in  Fig.~\ref{DipInt}(b). The wave functions were determined using the tight-binding (TB) approach to be described in Sec.~\ref{Calculation_procedure}. This method enables the determination of the distortion of the wave functions induced by the applied electric field, which  makes the dipole moments field-dependent (Sec.~\ref{Molecular binding energy}). We have also applied the TB method to account for modifications of the potential in one DQW induced by the presence of an IX in the other DQW. We show that polarization effects induced by the interaction between two IXs lead to non-negligible corrections to the binding energy of IX molecules as well as of indirect trions, as is described in detail in Sec.~\ref{Indirect exciton-carrier interactions}.

The large inter-layer binding energies reported in Ref.~\onlinecite{PVS315} were measured in an IX density regime, where the average separation between IXs within a DQW-bilayer is comparable to the separation between the bilayers, thus indicating that they may arise from many-body interaction effects, which are addressed in Sec.~\ref{Many-particle effects}. Here, we start by considering the coupling between a single dipole in one DQW bilayer with a cloud of dipoles in the adjacent DQW (Sec.~\ref{Indirect exciton lattices}). The dipoles in the cloud are assumed to be arranged  in a regular lattice. We show using an electrostatic model  that the lattice dipolar interaction energy reduces with the lattice density. This model is complemented by an exact diagonalization model (Sec.~\ref{Exact-diagonalization study of IXs}) for the interaction between one-dimensional (1D) dipolar bilayers, which predicts the experimentally observed minimum of the binding energy for equal IX densities in the two DQW bilayers.  The IX inter-layer binding energies predicted by the lattices models are, however, much smaller than the measured ones.

The last sections are devoted to \aPVS{di-}polaron models for the interlayer interaction. We first demonstrate that the local deformation of the lattice due to the coupling to a single IX in a remote DQW can lead to the formation of a \aPVS{di-}polaron, which further  increases the IX binding energy (Sec.~\ref{IX Lattice Polaron}). The lattice \aPVS{di-}polaron corrections to the inter-layer binding energies are, however, very small.
Finally, Sec.~\ref{Continuous Polaron} introduces a continuous model for a Fr{\"o}hlich-type  \aPVS{di-}polarons based on an interaction Hamiltonian, which can be exactly diagonalized. An analogous treatment has been recently proposed for dipolar cold atoms~\cite{Pena_arXiv_1804_06390}. Preliminary results for \aPVS{di-}polarons in coupled DQW bilayers have been presented in Ref.~\onlinecite{PVS315}. This model allows us to address the impact of interactions within the IX fluids on the \aPVS{di-}polaron formation energy.  Its predictions depend on the correlation state of the IX gas. In the low density regime, where correlations are not expected to be important, the interlayer binding energy increases with the gas density. As the density increases, one expects a transition to a correlated regime, where the interlayer binding energy is small and reduces with density. This behavior is in qualitative agreement with the predictions of the lattice model, which implicitly assumes correlated species. The \aPVS{di-}polaron model also semi-quantitatively describes the density dependence of the inter-layer binding energies reported in  Ref.~\onlinecite{PVS315}.
The main conclusions are summarized in Sec.~\ref{Conclusions}. 

\section{Calculation procedure}
\label{Calculation_procedure}

We have investigated the prototype structure sketched in Fig.~\ref{DipInt} consisting of two DQWs, each formed by two 10~nm-wide (001) GaAs QWs separated by 4~nm-wide outer Al$_{0.3}$Ga$_{0.7}$As barriers. The DQWs are separated by a 10~nm-thick~Al$_{0.3}$Ga$_{0.7}$As barrier thus yielding a  vertical separation between the DQW centers of  $L_z=34$~nm [cf.~Fig.~\ref{DipInt}(b)].  

Equation~\ref{EqUdd} yields the electrostatic interaction energy between two IXs in the PCA in the far field. Since  the vertical separation between the dipoles is only a few times larger than the dipole length $d_l$ for the studied  sample, we have used for the determination of the interactions the full expression for the electrostatic energy between four charged particles given by: 

\begin{widetext}
\begin{equation}
U^{(c)}_{dd}(\textbf{r}) =\frac{e}{4\pi \varepsilon\varepsilon_0} \left(-\frac{1}{\sqrt{(d_l+L_z)^2+r_\mathrm{\parallel}^2}}-\frac{1}{\sqrt{(L_z-d_l)^2+r_\mathrm{\parallel}^2}}+\frac{2}{\sqrt{L_z^2+r_\mathrm{\parallel}^2}}\right).
\label{EqUddc}
\end{equation}
\end{widetext}

\noindent Here,  $e$ is the electron charge,  $\epsilon=12.9$ the effective dielectric constant of the layer structure, and $d_l=p/e$ the length of the dipoles. The first two terms within the parenthesis arise from the attractive interaction between the IX charges of opposite signs in DQW$_1$ and DQW$_2$, while the last one accounts for the repulsion between charges of the same sign. 

The electronic structure of the IXs was determined using the  empirical tight-binding (TB) method.\cite{Slater54a} The  simplicity of the TB method enables the microscopic determination of the band structure of DQWs with QW widths up to a few tens of nm using moderate computational efforts.
Although empirical in the sense that it uses parameters fitted to the bulk band structure, it can be regarded as ``microscopic'' when compared with other approaches such as the  {$\bf {k\cdot p}$} effective mass calculations. 
The TB calculations used a basis consisting of  $sp^3s^{*}$ orbitals~\cite{Vogl83a,Vogl_JPCS44_365_83} including spin-orbit coupling\cite{Hass83a,Chadi77a}. This orbital basis consists of 10  orbitals per atom and includes only nearest-neighbor interaction.  It has been shown to reproduce well the highest valence 
and the lowest conduction bands of most bulk semiconductors.\cite{Vogl83a} 

%
%
\begin{table}
\caption{ Tight-binding parameters (in units of eV) used in the calculation of IX interactions. 
The notation used correspond to the one of Ref. \protect\onlinecite{Lok93a}.
In order to account for the valence band discontinuity we subtracted 
0.16~eV from the diagonal parameters (i.e., $E_s$, $E_p$, and 
$E_\mathrm{s^*}$) of the Al$_\mathrm{0.3}$Ga$_\mathrm{0.7}$As layers. 
 }
\begin{tabular}{| l | c | c |}
\hline
Parameter 	&  GaAs\cite{Schulman85a} (eV)	& Al$_\mathrm{0.3}$Ga$_\mathrm{0.7}$As
(eV)\\
\hline
$E_s$(anion) 	&	-8.457	&	-8.1774	\\
$E_p$ (anion) 	&	0.9275	&	0.93125	\\
$E_s$(cation) 	&	-2.7788	&	-2.29466	\\
$E_p$ (cation) 	&	3.5547	&	3.55029	\\
$V_\mathrm{ss}$ 	&	-6.4513	&	-6.51511	\\
$V_\mathrm{xx}$ 	&	1.9546	&	1.94782	\\
$V_\mathrm{xy}$ 	&	4.77	&	4.6122	\\
$V_\mathrm{sp}$(anion,cation) 	&	4.48	&	4.6672	\\
$V_\mathrm{sp}$(cation,anion) 	&	7.85	&	7.1378	\\
$E_\mathrm{s^*}$(anion) 	&	8.4775	&	8.17915	\\
$V_\mathrm{s^*p}$(anion,cation) &	4.8422	&	4.73714	\\
$E_\mathrm{s^*}$(cation) 	&	6.6247	&	6.65539	\\
$V_\mathrm{s^*p}$(cation,anion) &	7	&	6.3988	\\
$3\lambda_a$ (anion) 	&	0.39	&	0.39342	\\
$3\lambda_c$ (cation) 	&	0.174	&	0.129	\\
\hline
\end{tabular}
\label{TBTable}
\end{table}

The TB calculations were carried out for a periodic super-cell  consisting of a single DQW  sandwiched between two 8~nm wide Al$_\mathrm{0.3}$Ga$_\mathrm{0.7}$As outer barriers. The tight-binding parameters employed  for the GaAs and Al$_\mathrm{0.3}$Ga$_\mathrm{0.7}$As layers are summarized in Table~\ref{TBTable}. The effects of the  external electric field $F_z$ were taken into account by adding the position-dependent electrostatic energy to the on-site TB parameters. A similar procedure was used to address IX-IX interactions, as will be described in detail in the next sections.

\begin{figure}[!tbhp]
\begin{center}
\includegraphics[width=.95\columnwidth,angle=0, clip]{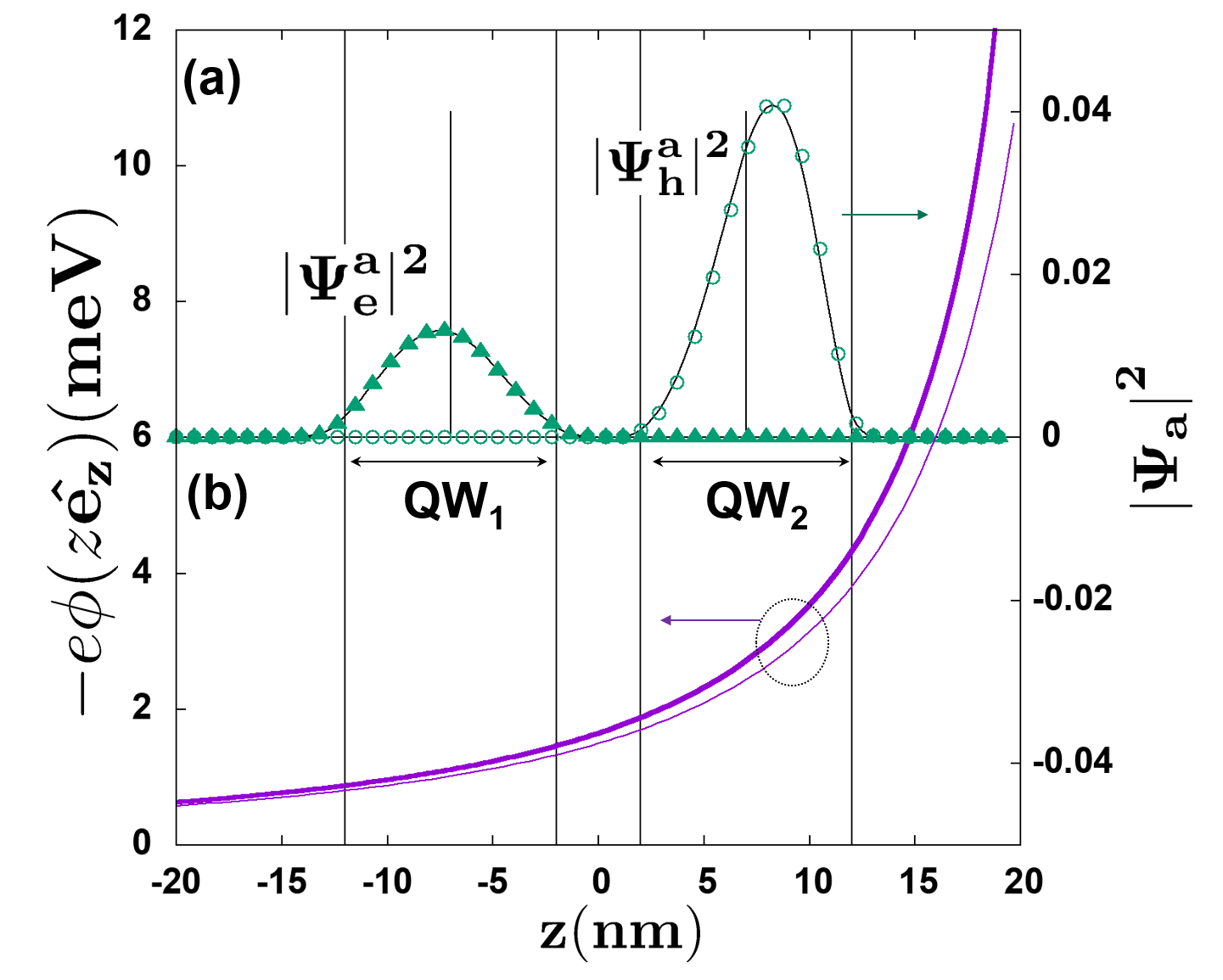} 
\end{center}
\caption{
(a) Squared tight-binding wave functions projected on the anion sites (superscript $a$) for the lowest conduction ($|\Psi^a_e|^2$) and highest valence  band state $|\Psi^a_h|^2$ calculated for an applied field $F_z=20$~kV/cm (thin lines). The symbols show the corresponding wave functions obtained using the self-consistent approach to take into account for the  presence of an IX in the adjacent DQW (see text for details).
(b) Spatial dependence of the electronic potential energy [-$e\phi(z)$] generated by the IX in DQW$_2$ on the atomic sites of DQW$_1$ [cf.~Fig.~\ref{DipInt}(b)] for zero lateral separation $x$ as determined using the self-consistent TB approach. The thin line  shows the corresponding potential for DQW$_2$ calculated in the point charge approximation.
}
\label{FigWave}
\end{figure}

\section{Single IX Interactions}
\label{Single IX Interactions}

\subsection{Exciton-exciton interactions}
\label{Exciton-exciton interactions}

The thin line in Fig.~\ref{FigWave} displays the spatial dependence of the TB wave functions of the lowest conduction band state $|\Psi^a_e|^2$ and highest valence  band state $|\Psi^a_h|^2$ calculated for a single DQW
under an applied field $F_z=20$~kV/cm. The electron and hole wave functions are normalized according to $\int_{-\infty}^{\infty}  {|\Psi^a_{e(h)}|^2}dz=1$.  
For clarity, we only show  the  projection of the wave function on the anion sites (superscript $a$). The larger amplitudes for holes than for electrons reflect the fact that the hole wave functions for the band-edge states have higher amplitudes on the anion sites (the opposite applies for the cation sites). The applied electric  field makes the electron and the hole wave functions asymmetric with a center of mass shifted with respect to the center of the QWs. Note that the electron and hole wave functions shift  in opposite directions under the applied field,  thus increasing the electric dipole moment of the IX species.

\begin{figure}[!tbhp]
\begin{center}
\includegraphics[width=1\columnwidth,angle=0, clip]{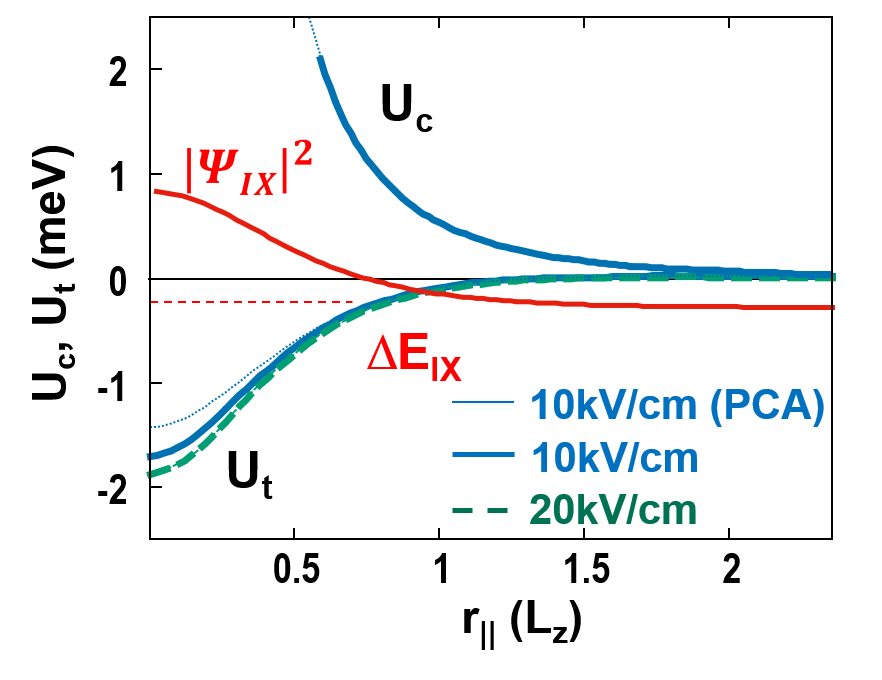} 
\end{center}
\caption{Interaction energy between two IXs in the same ($U_c$) and in different DQWs ($U_t$) as a function of  their lateral separation (in units of $L_z=34$~nm). The calculations for $U_t$ were carried out for electric fields $F_z$ of 10 and 20 kV/cm. The thin line displays $U_t$ as determined in the point charge approximation of  Ref.~\onlinecite{PVS284}. The horizontal dashed line marks the binding energy of the single IX molecular state confined in this potential, $\Delta E_\mathrm{IX}$ (calculated for $F_z=10$~kV/cm). The corresponding wave function amplitude $|\Psi|^2$ is also displayed.
}
\label{FigU}
\end{figure}

The interaction energy between two IXs [IX$_1$ and IX$_2$  located in  DQW$_1$ and DQW$_2$, respectively, cf.~Fig.~\ref{DipInt}(b)], was calculated  using the following self-consistent approach. First, the TB wave functions calculated  for one DQW (assumed to be DQW$_2$) were used to determine the charge distribution along the z-direction in the location of the remote DQW (DQW$_1$). The latter was then applied to  correct the on-site electronic potentials ($\Delta\phi(r_\mathrm{\parallel},z_\mathrm{i,1})$) on  the neighboring DQW according to:  
 
\begin{equation}
\Delta \phi(r_\mathrm{\parallel},z_\mathrm{i,1}) = \frac{e}{4\pi \epsilon\epsilon_0} \sum_\mathrm{z_\mathrm{j,2}}
\left[ 
\frac{|\Psi_\mathrm{h,2}(z_j,2)|^2 - |\Psi_\mathrm{e,2}(z_j,2)|^2}{ \sqrt{r_\mathrm{\parallel}^2+(L_z + (z_\mathrm{j,2}-z_\mathrm{i,1}))^2}}
\right]. 
\label{EqP}
\end{equation}

\noindent Here, the subscripts $1$ and $2$ denote, respectively, the coordinates across the IX structures in  DQW$_1$ and DQW$_2$. 
The energy eigenvalues and the wave functions for DQW$_1$ were then calculated under the corrected potential. This procedure was then repeated until self-consistency was reached, as defined by changes in   $\Delta\phi$ in successive iteration steps of less than $10^{-5}$~eV.  In general, the procedure converges within a few iteration steps.

The symbols in Fig.~\ref{FigWave} show the self-consistent electron and hole wave functions for two coupled DQWs under the  interaction potential given by Eq.~(\ref{EqP}). The shape of the  wave functions is very similar to the ones for IXs in an uncoupled DQW (thin line), except for the slightly larger shifts with respect to the center of the QWs (the shifts are smaller than the size of the symbols for $|\Psi^a_{e(h)}|^2$). The latter translates into a slightly enhanced IX electric dipole moment. The thick solid line in the lower panel of the figure displays the spatial dependence of the self-consistent potential  $ \phi_e(x=0,z_\mathrm{i,1})$. 
 The dashed line shows, for comparison, the dipole potential calculated in the PCA. In both cases, the electronic potential increases towards positive $z$ due the presence of the second dipole centered at $z=\ell_z$ [cf.~Fig.~\ref{DipInt}(b)].  The slightly larger values for the self-consistent potential arise from the extended character of the wave functions as well as from the distortions of the  wave functions induced by the remote IX dipole. 

The potential interaction energies $ U_\mathrm{c} (x, L_z=0)$ and  $ U_\mathrm{t} (x, L_z)$ for an IX molecule as well as for two IXs in the same DQW, respectively, can be determined as the difference between the self-consistent TB transition energies and the one for an isolated IX. The dependence of  these energies on lateral separation $r_{\parallel}$  is plotted in Fig.~\ref{FigU}. As expected, the intra-DQW interaction is always repulsive and stronger in magnitude than the inter-DQW one.
The thin line represents for comparison the  dipolar energy  determined in the PCA (cf.~Ref.~\onlinecite{PVS284}) for an electric field $F_z=10$~kV/cm. As expected, the minimum energy configuration corresponds to two stacked IXs (i.e., with lateral separation  $r_{\perp}=0$). 

The self-consistent interaction energy $U_t$  is approximately 20\% larger than the corresponding one  ($\Delta U_\mathrm{t,PCA}$) from the PCA.   This inter-DQW polarization behavior can be understood with the help of Fig.~\ref{FigWave}. In the PCA, the interaction energy  $U_\mathrm{t,PCA}$ is simply given by the difference between the electrostatic potentials $|-e\phi_\mathrm{t,PCA}|$ (thin line) at the center of the two QWs. The self-consistent electron and hole TB wave functions, in contrast, extend over the width of the QWs and also penetrate the barriers.  The deformation of the wave functions together with the strong  z-dependence of  $-e\Delta \phi (x,z)$  ($\approx r^{-3}$)  shift the energies of the electron and hole eigenstates towards opposite directions, thus increasing the IX binding energy.

\begin{figure}[!thbp]
\begin{center}
\includegraphics[width=.95\columnwidth,angle=0, clip]{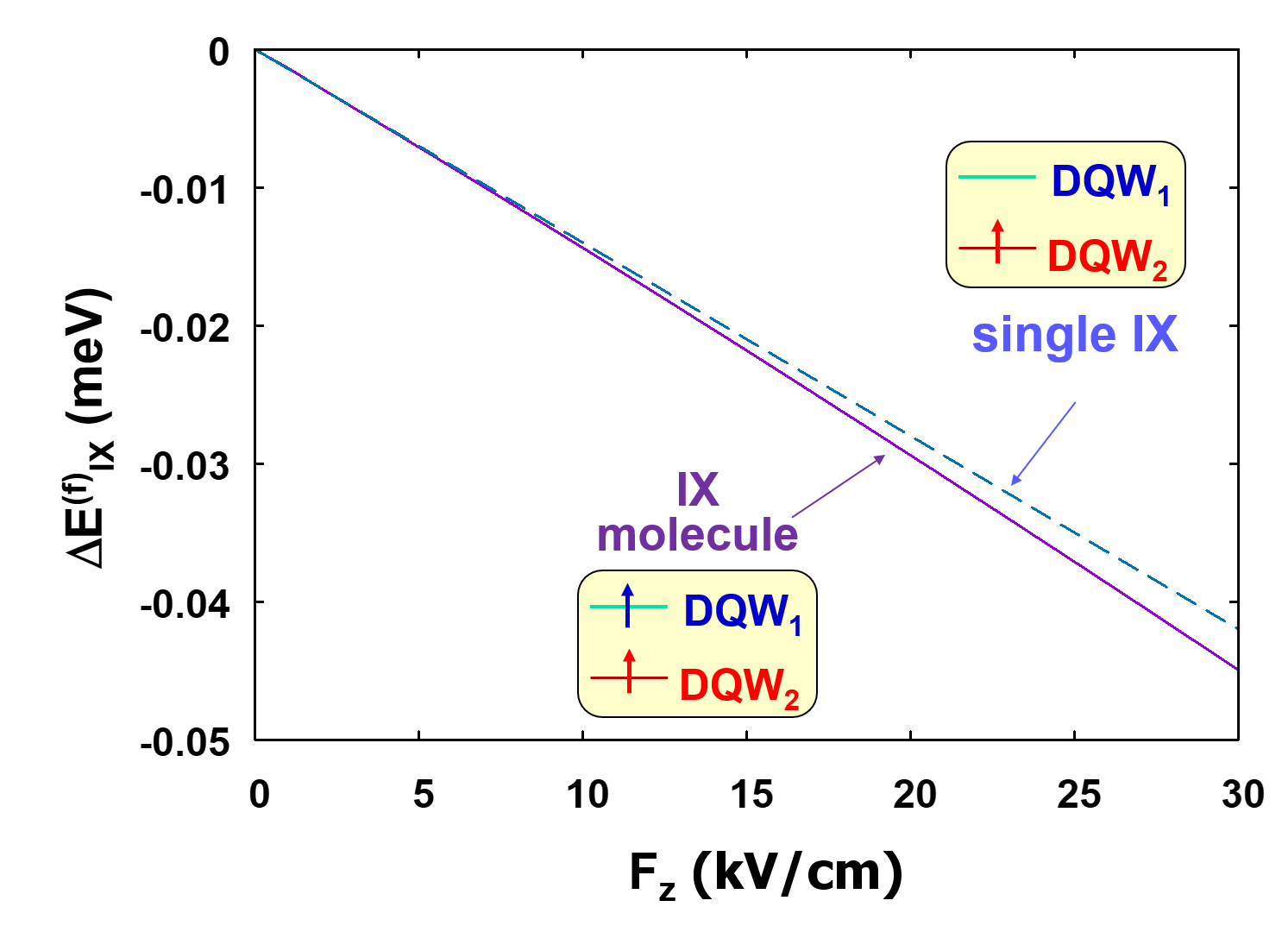} 
\end{center}
\caption{Electric-field-induced changes $\Delta E^\mathrm{(f)}_\mathrm{IX}$ of the resonance energy of a single IX (dashed line) and of an IX-IX molecule (solid line).
}
\label{FigField}
\end{figure}

The deformation of the wave functions makes the IX dipole moment dependent on the applied field.  The impact of the field polarizability  becomes evident when one compares the field-induced changes $\Delta E^\mathrm{(f)}_IX$ of the resonance energy of an IX-IX molecule with the one of an isolated IX, as illustrated, respectively, by the solid and dashed lines in Fig.~\ref{FigField}.

\subsection{Molecular binding energy}
\label{Molecular binding energy}

The attractive inter-DQW potential $U_t$ of Fig.~\ref{FigU} can bind two IXs to form an IX molecule. 
In order to obtain the molecular binding energy, we numerically solved\cite{Arsoskia_arXiV_15} the Schr{\"o}dinger equation for the relative motion of an IX-IX pair under the radial potential $U_t(r)$. The calculations were carried out assuming a reduced effective mass for the IX molecule equal to the combined mass of the electron ($m_e$) and heavy-hole ($m_{hh}$) in GaAs, i.e.,  $\frac{1}{2}(m_e+m_\mathrm{hh})$ with $m_e=0.067$ and $m_\mathrm{hh}=0.22$. In agreement with Ref.~\onlinecite{PVS284}, we found that the inter-DQW potential supports a single bound state with zero angular momentum and energy $\Delta E_\mathrm{IX}$ and wave function $|\Psi_\mathrm{IX}|^2(r)$ displayed in Fig.~\ref{FigU}. It is interesting to note that the strong kinetic corrections due to the small reduced mass and short extension of the attractive potential yields a  binding energy $\Delta E_\mathrm{IX}=0.28$~meV much smaller than the potential depth $U_t(r_{\parallel}=0)=-1.7$~meV determined for $F_z=10$~kV/cm.

The shape of the  inter-DQW potential $U_t(r)$  may be affected by lateral (i.e., along the DQW plane) distortions of the electron and hole wave functions induced by IX-IX interactions, which cannot be addressed using the 1D tight-binding approach employed here. 
We have estimated the impact of these distortions in the PCA by assuming that the electron and the hole charges can be displaced by small amounts $-\Delta r$ and $\Delta r$, respectively, in order to minimize the electrostatic interaction energy between two IXs (note that such a distortion does not change the electrostatic energy associated with the vertical field $F_z$). We found that $U_t(\Delta r, r_0)$ can indeed be minimized by a non-vanishing $\Delta r$. The energy reduction (on the order of a few tens of $\mu$eV) and magnitude of the shifts ($|\Delta r|<0.5$~nm), however, are too small to introduce sizable corrections to $U_t(r)$ or to the binding energy.

\begin{figure}[tbhp]
\begin{center}
\includegraphics[width=0.95\columnwidth,angle=0, clip]{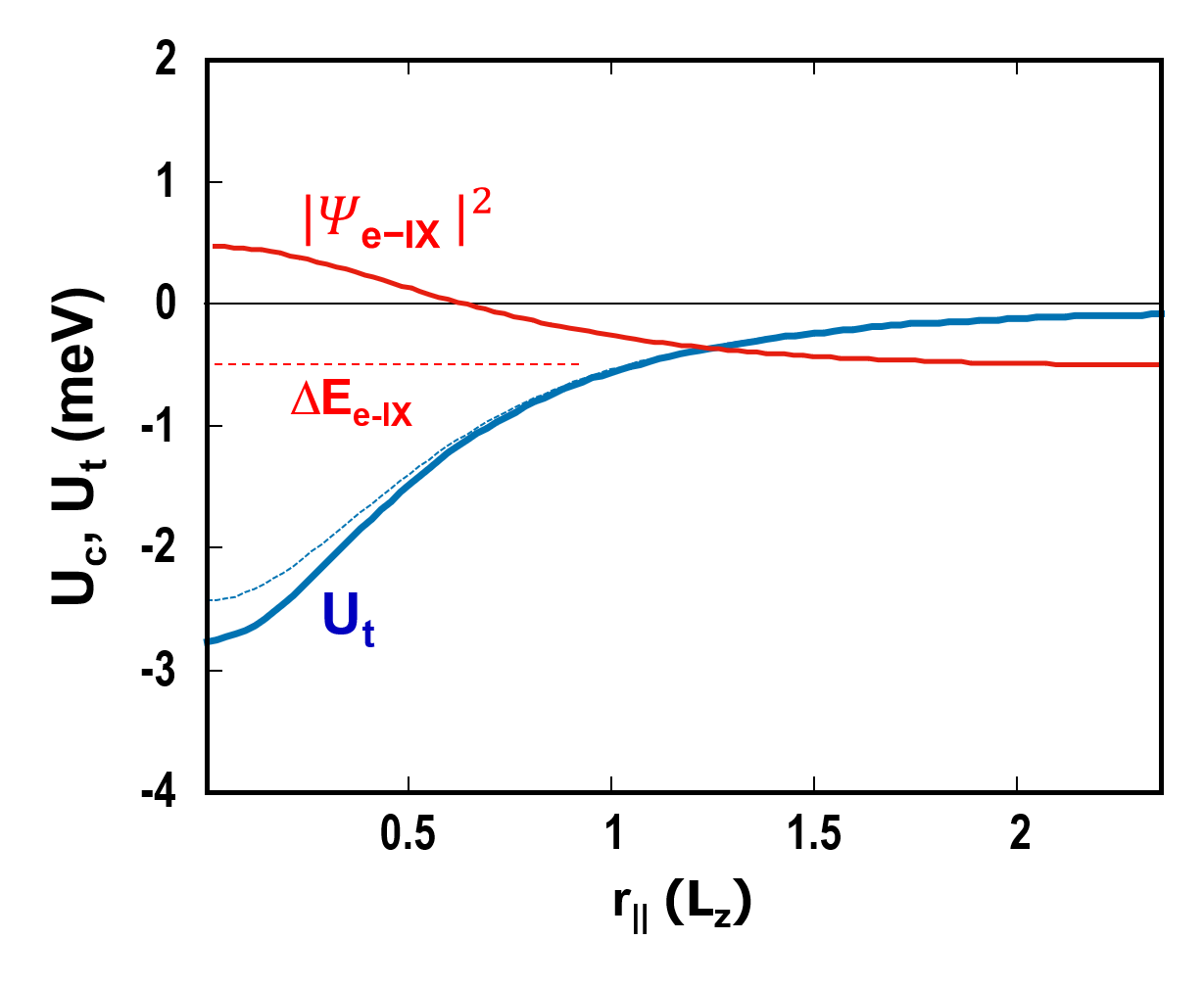} 
\end{center}
\caption{
Interaction energy $U_t$ between an indirect trion $e$-IX consisting of an IX in one DQW and an electron in the adjacent DQWs as a function of the lateral separation $r_{\parallel}$ calculated for    $F_z=10$~ kV/cm. The dashed line displays $U_t$ determined in the point charge approximation of  Ref.~\onlinecite{PVS284}. The horizontal dashed line marks the energy of the single confined state in this potential, $\Delta E_\mathrm{e-IX}$. The corresponding wave function amplitude $|\Psi_\mathrm{e-IX}|^2$ is also displayed.   
}
\label{FigTrion}
\end{figure}

\subsection{Indirect exciton-carrier interactions}
\label{Indirect exciton-carrier interactions}

IXs in one DQW can also couple to an electron or a hole in the other DQW to form a spatially indirect trion. We have extended the  TB approach to estimate the  binding energy of these species. The procedure is essentially the same as the one used in the previous section, except that we neglect the hole charge distribution [e.g., $\Phi_\mathrm{h,2}$ in Eq.~(\ref{EqP})]. The results for the interaction of an IX in DQW$_1$ with an electron in the lower QW of DQW$_2$ (cf.~Fig.~\ref{DipInt}) are summarized in Fig.~\ref{FigTrion}.  When compared to the inter-DQW attractive potential $U_t(r)$ between two IXs  of Fig.~\ref{FigU}, the electron-IX interaction is approximately 50\% stronger has  also a longer lateral decay length. This potential supports a bound state with a binding energy $\Delta E_\mathrm{e-IX}=0.53$~meV  almost twice as large as the one for an IX molecule. 
Similar results apply to hole trions. The interaction of trions with the electric field is also increased by approximately 33\%.

\section{Many-particle effects}
\label{Many-particle effects}

The inter-DQW binding energy for the remote interactions between two IXs determined in the previous section are larger compared to the values obtained in the PCA but still much smaller than the one measured in Ref.~\onlinecite{PVS315}. This discrepancy has motivated us to analyze dipolar interaction effects involving many particles.

\begin{figure*}[!thbp]
\includegraphics[width =.95\textwidth]{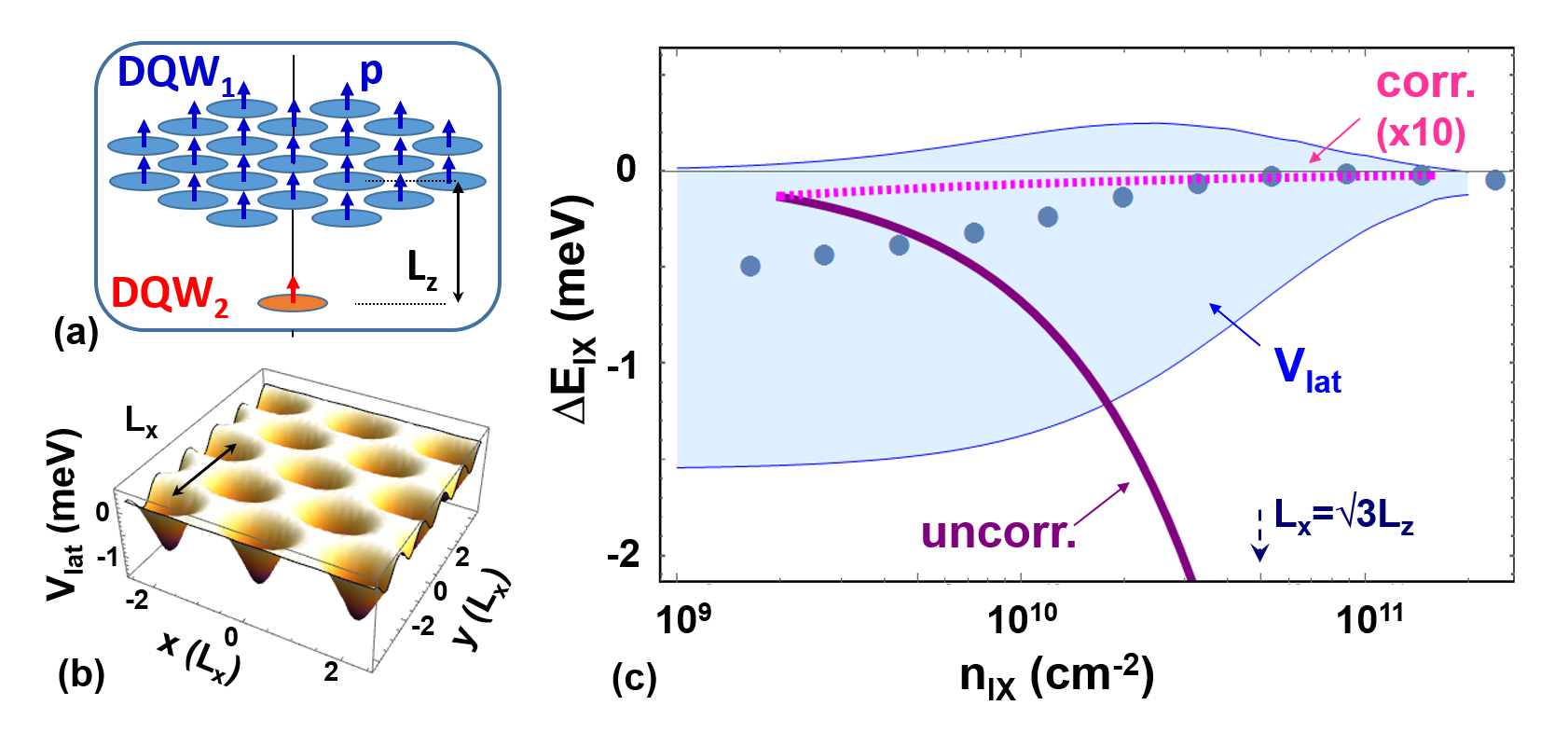}
\caption{ 
(a) Interaction between a single IX in DQW$_2$ with a closed-packed IX lattice in DQW$_1$. 
(b) Dipolar potential on the single IX in DQW$_2$ induced by the  lattice in DQW$_1$ with  a lattice constants  $L_x=2 L_z $ and density $n_\mathrm{IX}={2}/({\sqrt{3}L_x^2})$. 
(c) Shaded region: range of the local dipolar interaction potential [$V_\mathrm{lat}$, cf. Eq.~(\ref{EqIXP1})] experienced by the single IX in (a) as a function of lattice density.  The superimposed dots are the lowest energy IX quantum state in this potential, which correspond to the IX binding energy ($\Delta E_\mathrm{IX}$). 
The vertical dashed arrow marks the density for which $L_x=\sqrt{3}L_z$. 
The thick dotted and solid lines display the prediction of the continuous \aPVS{di-}polaron model in the limit of fully correlated [corr., cf. Eq.~(\ref{E02})] and uncorrelated [uncorr., cf. Eq.~(\ref{E03})] IX gases, respectively.
}
\label{fig:Theory}
\end{figure*}

\subsection{Indirect exciton lattices}
\label{Indirect exciton lattices}

The previous results for the inter-DQW dipolar interaction apply for low IX densities, i.e., when the average separation $L_{x}$ between  IX within a DQW far exceeds the distance $L_{z}$  between the DQWs. For larger densities,  multi-particle interaction effects modify the  inter-DQW interaction potential with respect to the simple form given by Eq.~(\ref{EqUdd}). In order to make the many-body problem tractable, we first neglect kinetic effects and assume that the high-density IXs are arranged in a regular lattice. 
We note that since dipolar interactions fall rather quickly with increasing distance, mostly contributions of nearest and next-nearest neighbors  are important. This means that ordered lattice models  also approximate a correlated liquid with only short-range order~\cite{Laikhtman_PRB80_195313_09,Mazuz-Harpaz_PNAS116_18328_19}.
A stable lattice configuration requires an isometric lattice (e.g., a square, triangular or hexagonal lattice). We have opted for a triangular lattice, since it provides the maximum number of nearest neighbors.\cite{Remeika_CLEO_2_16a}
%
Specifically, we have considered the configuration illustrated in Fig.~\ref{fig:Theory}(a), where a single IX in DQW$_\mathrm{2}$ interacts with a triangular lattice of IXs generated by the lattice vectors ${\bf v_a} = (1,0) L_x$ and ${\bf v_b} = (\sqrt{3}/2,1/2) L_x$, where $L_x$ is the lattice constant.  
The particle density $n_{IX}$ in the triangular lattice is related to the lattice constant $L_\mathrm{z}$ according to $n_\mathrm{IX}={2}/({\sqrt{3}L_x^2})$.

The dipolar potential V\textsubscript{lat} experienced by the single IX can be obtained by assuming that the electrons and holes forming the dipoles in the lattice to be elementary point charges located at the center of the respective QWs in the stacked DQW structure. For an IX at position ${\bf r}=({\bf r_\mathrm{\parallel}}, r_\mathrm{\perp})$, where ${\bf r_\mathrm{\parallel}}$ and $\bf r_\mathrm{\perp}$ are the in-plane and out-of-plane displacement components, respectively, the potential is calculated  by summing the  $U_\mathrm{dd}({\bf r})$  contributions [cf. Eq.~(\ref{EqUddc})] summed over the lattice sites. 
The potential $V_\mathrm{lat}(\bf{r})$ experienced by the single IX in DQW$_2$ then becomes:

\begin{equation}
V_\mathrm{lat}(\bf{r_{\parallel}}, \bf{r_\mathrm{\perp}}) = \sum_\mathrm{i,j}^{\bf{r_{\parallel}}<r_\mathrm{max}}
U^{(c)}_\mathrm{dd}
\left[
\bf{r_\mathrm{\perp}}-  \left| i {\bf v_a} + j {\bf v_b} + \bf{r_\mathrm{\parallel}}\right|
\right]
\label{EqIXP1}
\end{equation}

\noindent Due to the extended nature of the potential, the summation needs to be carried out over a large number  of lattice around the central IX in order to ensure convergence. In order to reduce the computational efforts, the coupling to lattice IXs at lateral distances $r_{\parallel}>r_\mathrm{max}\approx 20~L_x$ was taken into account by replacing the summation in Eq.~(\ref{EqIXP1}) by an integration over  a continuous density of dipoles.

Figure~\ref{fig:Theory}(b) displays the potential experienced by the single IX as it moves underneath a lattice with lattice constant 
$L_\mathrm{x}=2 L_\mathrm{z}$. For  $L_\mathrm{x}>>L_\mathrm{z}$, the  potential $V_\mathrm{lat}$ around each site resembles the  one for the dipolar potential $U_\mathrm{dd}$  with the  minima of the potential  aligned with the lattice sites. The shaded region in Fig.~\ref{fig:Theory}(c) marks the energy range spanned by the lattice potential $V_\mathrm{lat}(\bf{r})$.
In the opposite limit $L_x << L_z$,  $V_{lat}({\bf\vec{r}})\rightarrow 0$, thus reproducing the well-known result that the electric field generated by an infinite sheet of dipoles vanishes at large distances. The  minima of $V_{lat}({\bf\vec{r}})$ are always higher than 
the potential  $U_{t}(0)$ for a single IX-IX pair [cf. Fig.~\ref{FigU}]
 - this model thus predicts a reduction of the IX binding energy with increasing lattice density. The last behavior arises from the dominating contribution of repulsive intra-DQW interactions with increasing densities.

 The quantum corrections required for the determination of the binding energy of the single IX to the lattice were determined by numerically solving the two-dimensional  Schr{\"o}dinger equation for a IX in the periodic lattice potential.
The dots in Fig.~\ref{fig:Theory}(c) display the density dependence of the  lowest energy level $\Delta E_\mathrm{IX}$. In the limit of small densities, the  binding energies $|\Delta E_\mathrm{IX}|$ are larger than the ones for an IX molecule (cf.~Fig.~\ref{FigU}). The latter is due to the fact that the lattice is assumed to be static, so that the kinetic energy of the particle becomes equal to the IX mass (rather than to reduced mass of two IX, as in the case of an IX molecule). Note again that the absolute values for the binding energies are  smaller than the depth of the interaction potential.


\subsection{Exact-diagonalization study of IXs}
\label{Exact-diagonalization study of IXs}

In this section we analyze many-particle effects employing a numerical exact-diagonalization method. This method is quite powerful in capturing long-range correlations, although it suffers from small system size effects. In order to be able to consider significantly larger systems, in this section we limit the discussion to 1D system. Namely, we consider two double quantum wire structures 
with parameters similar to the full two DQWs case.

In order to make \aPVS{the} numerical approach feasible, we assume that the IXs occupy  sites of an artificial IX lattice with a minimum lattice constant. Physically, the latter is equivalent to have a cutoff in wavenumber and to disregard short wavelength processes.  The kinetic energy of the IXs is included as terms of nearest neighbor hopping. We construct the supercell of the size $L^T_x=nL_x$, where $L_x$ in this section denotes the lattice constant of the artificial lattice. In order to mimic infinite lattice and isolate bulk physics from edge effects, we apply periodic boundary conditions to the supercell. This modifies the form of the interaction potential given by \aPVS{Eq.~\ref{EqUdd}} of the main text, because now the IX interacts both with the IXs in the supercell, as well as with the images of the IXs in other supercells. The modified interaction has the form

\begin{equation}
U^{PBC}(x)=\sum_{s=-\infty}^{+\infty}\mu^2\frac{(1-3\cos^2\theta_s)}{r^3_s},
\label{PBCInteraction}
\end{equation}    

\noindent  where  

\begin{equation}
\mu = p/\sqrt{4 \pi \varepsilon \varepsilon_0}
\label{Eqmu}
\end{equation}

\noindent For the intra-layer interaction $\theta_s=\pi/2$ and \rPVS{$r^2_s=a^2+ x^2_s$}{$r^2_s=x^2_s$} \rPVS{whereas}{and} for inter-layer interaction $\cos\theta_s=L_z/r_s$ and $r^2_s=x^2_s+L^2_z$. Here $x_s=x+sL^T_x$. 
\rPVS{The cutoff length scale $a$  is introduced to remove the divergence of the intra-layer interaction  when two IXs occupy the  same site. }
{For intra-layer interaction when two IXs reside on the same site the interaction is diverging and we introduce cutoff scale. }
Introduction of the cutoff phenomenologically accounts  for the fact that dipolar interaction will be modified when the separation between IXs \rPVS{becomes}{is} comparable to the IX effective dipole length. While the modified interaction has a complex form in real space, the Fourier transform is less cumbersome

\begin{align}
U^{PBC}_\mathrm{intra}(k)&=2\mu^2\frac{|k|}{a}K_1\left(|k|a\right), \\
U^{PBC}_\mathrm{inter}(k)&=2\mu^2\left(\frac{|k|}{L_z}K_1\left(|k|L_z\right)-k^2K_2\left(|k|L_z\right)\right),
\end{align} 

\noindent where $K_i(x)$ is the modified Bessel function of the second kind. 

\begin{figure}[!thbp]
\includegraphics[width=8.5cm]{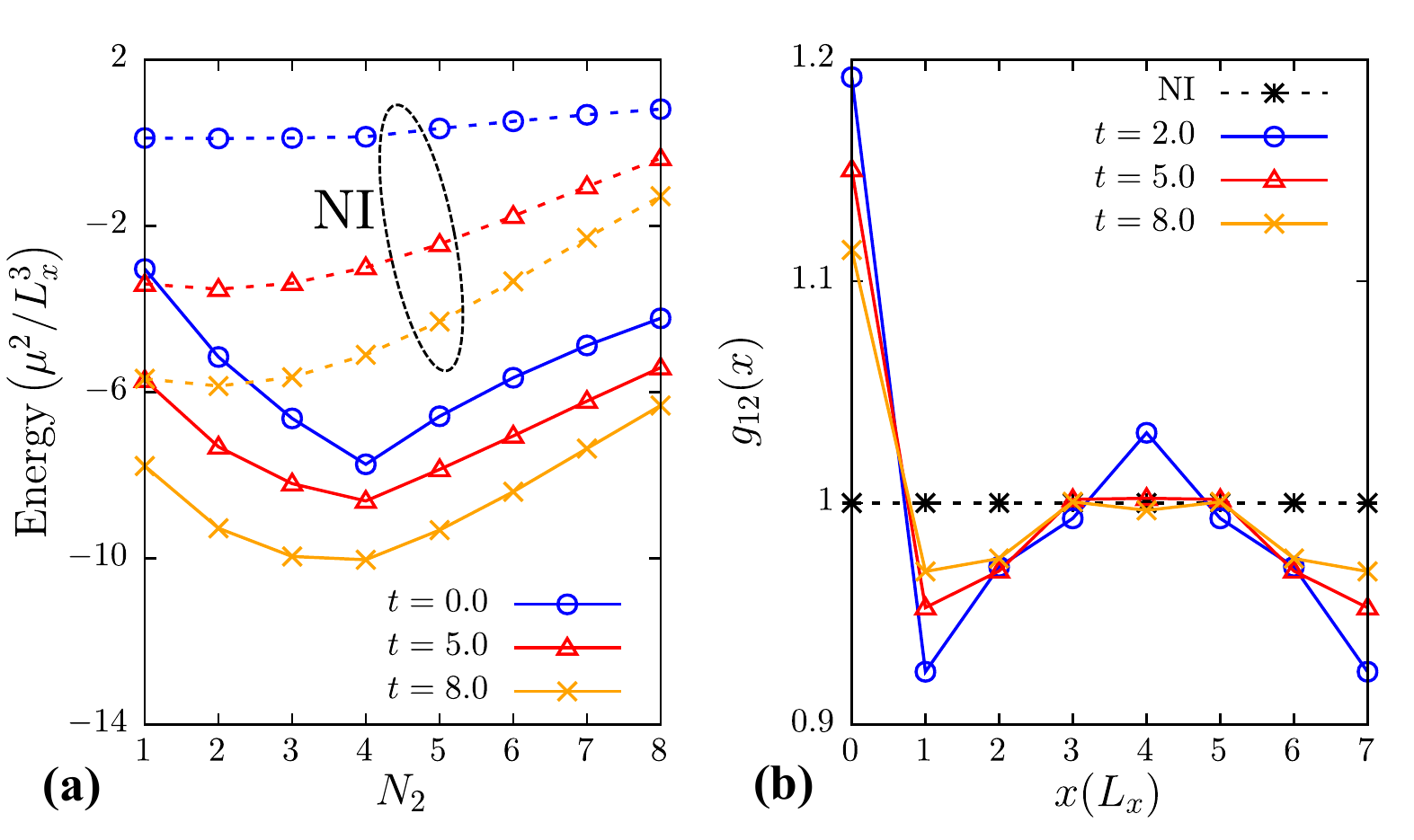}
\caption{\label{fig:EnergyPairCorr} (a) Dependence of the energy per IXs on number of IXs in the second layer when $N_1=4$ for different values of the hopping amplitude. (b) Inter-layer pair correlation function $g_{12}(x)$ for different values of hopping amplitude. Dashed lines in both figures correspond to the case when there is no inter-layer interaction (NI stands for no inter-layer interaction).   
}
\end{figure}

Once the Hamiltonian is formed we construct the full Hilbert space for a small number of particles and diagonalize the resulting Hamiltonian matrix. The diagonalization can be carried out both in real and momentum space. In real space the interaction expressed by Eq.~(\ref{PBCInteraction}) is diagonal, whereas in momentum space the kinetic term will be diagonal. We mostly work in momentum space, where the kinetic term takes the standard form $H_0(k)=t\cos\left(kL_x\right)$, where $t$ is the hopping amplitude. After diagonalization, the evaluated wave functions can be used to calculate physical properties of the ground state, such as pair correlation functions and  static structure factors. Using the translational invariance of the system the pair-correlation function can be stated in the form:

\begin{align}
g_{l^\prime l}(x)=\frac{1}{N_l^\prime N_{l}}\sum_{k_1,k_2,k_3,k_4}&\delta_{k_1+k_2,k_3+k_4}e^{i(k_1-k_2+k_3-k_4)\frac{x}{2}}\times \nonumber \\ 
&b^\dagger_{l^\prime k_1}b^\dagger_{l k_2}b_{l k_3}b_{l^\prime k_4},
\end{align}

\noindent where the index $l$ and $l^\prime$ denote the dipole layer (1 or 2), $N_l$ is the total number of IXs in layer $l$ and $b^\dagger_{l^\prime k}$, $b_{l^\prime k}$ denote, respectively,  the creation and annihilation operators for the state $k$ in layer $l$. Similarly,  for the structure factors we have

\begin{equation}
S_{ll^\prime}(q)=\frac{1}{N}\rho_l(q)\rho_{l^\prime}(-q)-\frac{N_lN_{l^\prime}}{N}\delta_{q0},
\end{equation}

\noindent where $q$ is the wave vector, $N=N_1+N_2$ is the total number of IXs and $\rho_l(q)=\sum_{k}b^\dagger_{lk}b_{lk-q}$ is the Fourier transform of the density operator.

\begin{figure}[!thbp]
\includegraphics[width=8.5cm]{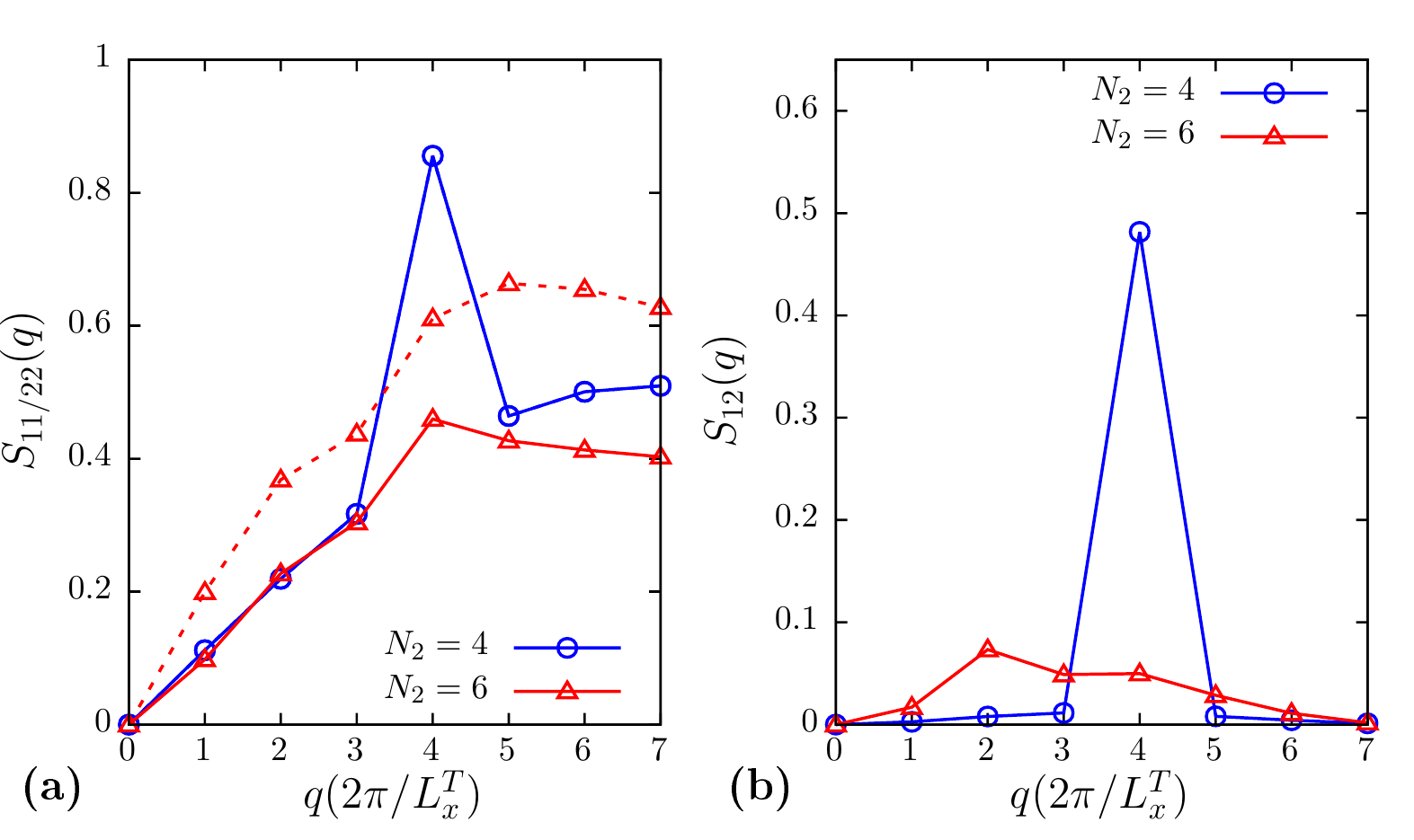}
\caption{\label{fig:StructureFactor} (a) Intra-layer structure factor for two cases of number of IXs in the second layer. Solid (dashed) lines correspond to $S_{11}(q)$ ($S_{22}(q)$). $S_{11}(q)$ and $S_{22}(q)$ are the same when $N_1=N_2$. (b) Inter-layer structure factor for two cases of number of IXs in the second layer. The hoping amplitude is set to $t=2.0$ in both figures.   
}
\end{figure}

The calculated  energies  will be presented in units of $\mu^2/L_x^2$ and the lengths will be normalized to $L_x$. The used parameters are $L_z=0.5$, $a=0.2$, $L^T_x=10$ which correspond to ones of the experiment in Ref.~\onlinecite{PVS315}. In one layer we fix the number of IXs to $N_1=4$ and vary the number of IXs in the other layer $N_2$ from 1 to 8. It should be noted, that due to the 1D character and small sizes of the  systems  considered here,  we will not be able to achieve quantitative agreement with the experiments. Our goal here is rather to gain a qualitative understanding of the physical characteristics of the system. In particular, we would like to understand the non monotonous dependence of the interlayer binding energy on the single layer IX density observed in Ref.~\onlinecite{PVS315}.

Figure~\ref{fig:EnergyPairCorr}(a) shows the dependence of the energy per IXs on the number of particles in the second layer $N_2$ for different values of the hopping amplitude. As  can be seen from the figure, the energy per IXs in the calculations  with inter-layer interaction has a minimum when $N_2=N_1=4$. This minimum is not present when inter-layer interaction is switched off. Addition of hopping deepens and smoothens out the minimum and, for large hopping amplitudes, can even shift the minimum towards smaller values of $N_2$. Similarly, the minimum can be pushed to larger values of $N_2$ by increasing $a$. This demonstrates the competition between three energy scales: the intra-layer interaction, the inter-layer interaction and the hopping energy. 

The energy minimum for $N_1=N_2$ in Fig.~\ref{fig:EnergyPairCorr} reproduces very well the one observed in the experimental results (cf.  Fig.~5 of Ref.~\onlinecite{PVS315}), which appears when the IX densities are approximately equal for both IX bilayers.
The appearance of the minimum energy in the experiments of Ref.~\onlinecite{PVS315} at $N_1\simeq N_2$ therefore suggests the following hierarchy of energy scales: the short-range intra-layer interaction is the largest energy scale when compared to inter-layer interaction, whereas the hopping  contribution is negligible as  compared to the other two energy scales. 
The IXs tend, therefore, to arrange themselves equidistantly in both bilayers by sitting on top of each other. The minimum energy is thus achieved when IX density is the same in both bilayers. 
%
The fact that the energies in Fig.~\ref{fig:EnergyPairCorr}(a) for cases with and without inter-layer interaction do not approach each other for large $N_2$ is  is a characteristic property 1D systems. Contrary to the 2D case, where the integrated inter-layer interaction between an IX and an infinite homogeneous distribution of IX vanishes, the interaction  between an IX and an infinite line of dipoles  is not zero. 
One has, therefore, a finite inter-layer contribution in 1D systems, which does not grow with $N_2$.

We next consider the inter-layer pair correlation function $g_{12}(x)$ [cf. Fig.~\ref{fig:EnergyPairCorr}(b)], which yields information about the distribution of IXs in the layer system. This figure clearly shows that when the inter-layer interaction is turned on the pair correlation function at $x=0$ becomes larger than one, which demonstrates that the probability of occupation of the same position in both layers is preferred. Increasing hopping amplitude reduces the peak value at $x=0$ essentially suppressing same site occupation in two layers, which again demonstrates the competition between the two energy scales.

Another way to understand the position of the minimum in Fig.~\ref{fig:EnergyPairCorr}(a)  is to consider the structure factor for the system. Figure~\ref{fig:StructureFactor} shows both the intra-layer and inter-layer structure factor for two values for the number of IXs in the second layer $N_2=4$ and $N_2=6$. For the case  $N_2=4$, both intra-layer and inter-layer structure factors show a peak when $q=4$, which signalizes the presence of short range order in the system. The fact that the peaks in the two plots appear for the same $q$'s  shows that intra- and inter-layer interactions are completely balancing each other and the system is its lowest energy state. When $N_2=6$, while first layer structure factors $S_\mathrm{11}$ and $S_\mathrm{22}$ still retain the peak at $q=4$,  the peak for the inter-layer structure factor $S_\mathrm{22}$ shifts towards smaller $q$'s. Therefore, in this case there is an imbalance between the two energy scales and intra-layer interaction determines the distribution of IXs in both layers. A similar suppression of the peaks in the structural factor as well as in the pair-correlation functions  is observed when the hopping amplitude is increased (not shown).       

In conclusion, the results of this section
 reproduce the experimentally observed  non-monotonous dependence of the IX energies on the one layer IX density shown in Ref.~\onlinecite{PVS315}. They also  show that the minimal energy per particle in stacked 1D bilayers is achieved for equal densities in both bilayers with the  IXs in one bilayer aligned with those in the other \added[id=AG]{one} to maximize the attractive component of the dipolar coupling.

\begin{figure}[!thbp]
\includegraphics[width=.97\columnwidth, keepaspectratio=true]{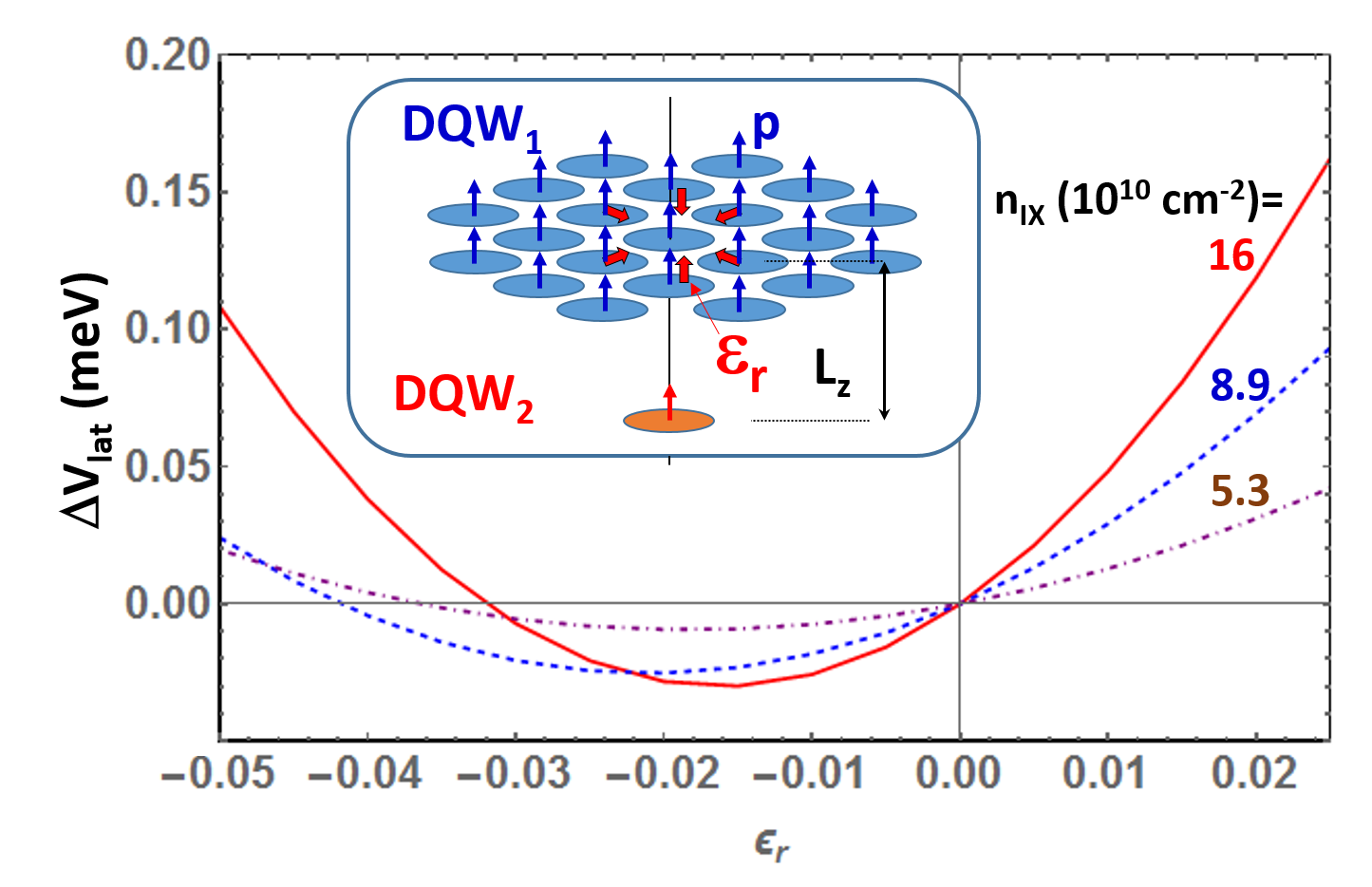}
\caption{
Dependence of the energy changes  
\ensuremath{\Delta}V\textsubscript{lat} of a single IX in DQW$_\mathrm{2}$ 
on the deformation \ensuremath{\epsilon}\textsubscript{r} of the closed-packed triangular 
lattice of IX in DQW$_\mathrm{1}$ displayed in the inset. The calculations were carried out for $L_\mathrm{z}=36$~nm.
and lattice constants $L_\mathrm{x}$ corresponding to IX densities $n_\mathrm{IX}={2}/({\sqrt{3}L_x^2})$ listed in the figure.
}
\label{FigIXP1}
\end{figure}

\subsection{IX \aPVS{di-}polarons}
\label{IX polarons}

The previous sections have addressed models for  interlayer excitonic interactions based on the arrangement  of the particles on fixed lattice sites. The interaction between the single IX in DQW$_\mathrm{2}$ and the lattice may also deform the IX lattice, thus leading to formation of an IX  \aPVS{di-}polaron  with increased interaction energy. This section discusses \aPVS{di-}polaron models for the dipolar interaction between IXs.

\subsubsection{IX Lattice \aPVS{Di-}polaron}
\label{IX Lattice Polaron}

We first consider a lattice \aPVS{di-}polaron arising from the distortion of the IX lattice in DQW$_1$ shown in Fig.~\ref{fig:Theory}(a). Due to the large values of the repulsive intra-DQW interaction in comparison with the attractive inter-DQW interaction, it is not obvious whether \aPVS{di-}polaron formation is energetically favorable. In 
fact, an IX molecule cannot bind an additional IX because the intra-DQW repulsive forces on the extra IX always exceed the attractive inter-DQW attraction [cf.~Fig.~\ref{DipInt}(c) and Ref.~\onlinecite{PVS284}].  In a locally deformed lattice, in contrast, the increased repulsion experienced by an IX when it approaches one of its neighbors is partially compensated by the reduction in repulsion as it moves away from the neighbors on the opposite side. This compensation mechanism enables the attractive inter-DQW to reduce the electrostatic energy and stabilize the \aPVS{di-}polaron. 

The \aPVS{di-}polaron formation energy  was calculated  for the triangular lattice shown in the inset of Fig.~\ref{FigIXP1}  by assuming that the neighbors of the central IX (i.e., the one aligned with the single IX in DQW$_2$) can displace radially by an amount $\epsilon_{\mathrm{r_{\parallel}}} L_\mathrm{x}$, where  $\epsilon_{\mathrm{r_{\parallel}}}$ is given by the Ansatz form: 

\begin{equation}
\epsilon(r_\mathrm{\parallel})=\epsilon_r e^{-\left[\frac{ r_\mathrm{\parallel} - L_\mathrm{x} }{r_\mathrm{\epsilon}}\right]},\quad r_\mathrm{\parallel}>0
\label{EqIXP2}
\end{equation}

\begin{figure}[!thbp]
\includegraphics[width=0.95\columnwidth, keepaspectratio=true]{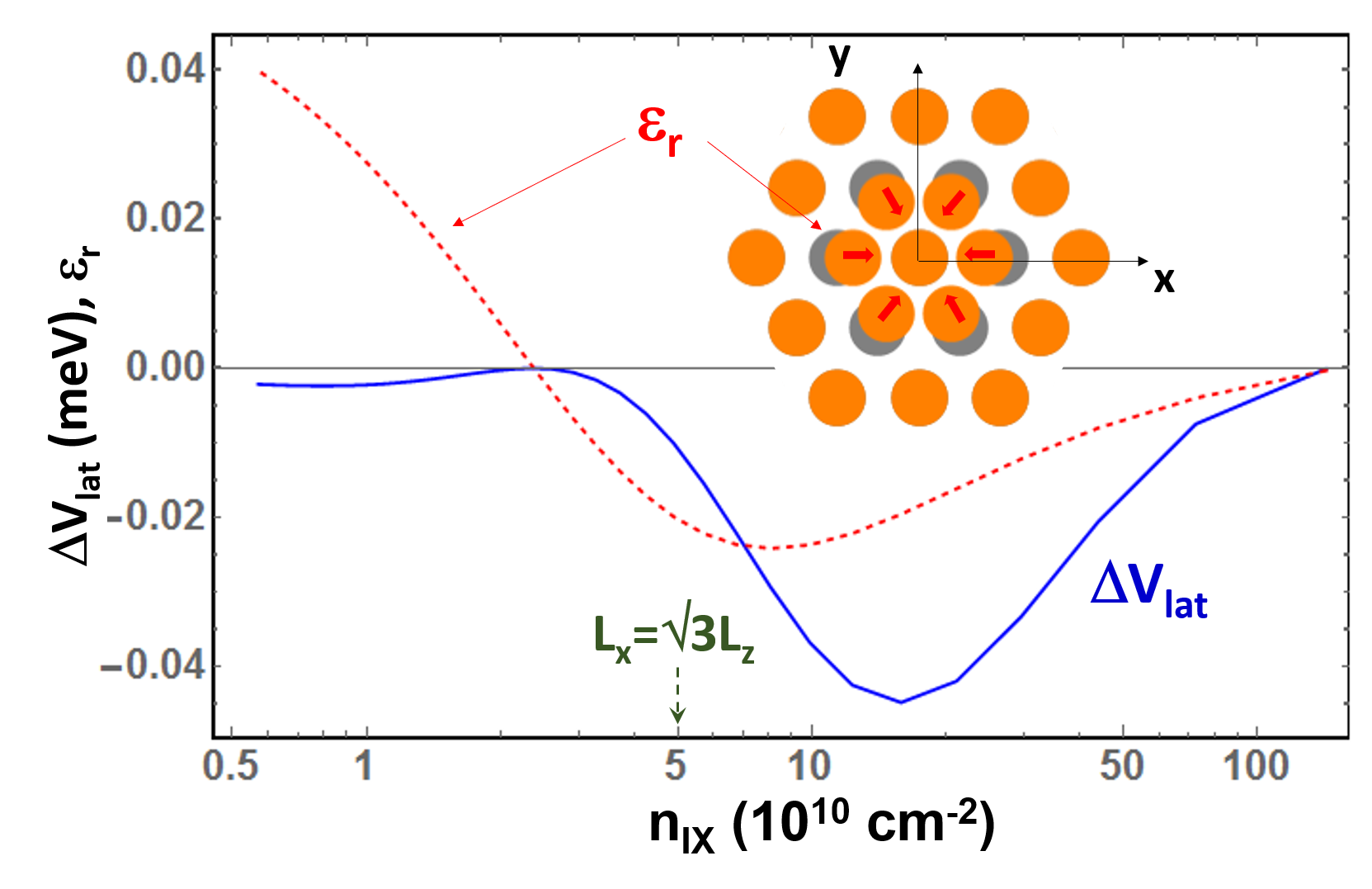}
\caption{ Equilibrium values  for the deformation ($\epsilon_\mathrm{r}$)  and corresponding change in potential ($\Delta V_\mathrm{lat}$) as a function of lattice density $n_\mathrm{IX}$. The orange 
circles in the inset show the deformation of the originally triangular IX lattice (gray circles)  due to \aPVS{di-}polaron formation around 
the central site bound to an IX in the adjacent DQW layer.}
\label{FigIXP2}
\end{figure}

\noindent Here, $\epsilon(r_\mathrm{\parallel})$  yields the relative radial displacement of the IXs at a distance $ r_\mathrm{\parallel}$ from the center of the  \aPVS{di-}polaron, $\epsilon_\mathrm{r} r_\mathrm{\parallel}$ is the lateral decay length of the deformation, and $\epsilon_\mathrm{r}$ the relative displacement of the first neighbors of the central IX located at $r=L_x$.
 According to the definition in Eq.~(\ref{EqIXP2}), $\epsilon_\mathrm{r}$ corresponds to the relative displacement of the nearest neighbors.

Figure~\ref{FigIXP1} shows the effect of the deformation on the potential experienced by the single IX (quantified in terms of the energy difference $\Delta V_\mathrm{lat} = V_\mathrm{lat}(0,0,\epsilon_r) -V_\mathrm{lat}(0,0,\epsilon_r=0)$ introduced by the deformation \ensuremath{\epsilon}\textsubscript{r}) as a function of the lattice density. The minima of \ensuremath{\Delta}V\textsubscript{lat}(\ensuremath{\epsilon}\textsubscript{r}) occur for $r_\mathrm{\epsilon}<<L_\mathrm{x}$, thus indicating that the deformation is restricted to the first neighbors. For this reason, the subsequent calculations were carried out 
assuming that the deformation is restricted to the first ring of IXs around the single IX in DQW$_2$. 
The plots show that the total energy reduces for small negative deformations (on the other of a few \%), which arise from the attraction of the neighbors by the single IX.

Figure~\ref{FigIXP2} displays the equilibrium values for  $\epsilon_\mathrm{r}$ and $\Delta V_\mathrm{lat}$ obtained 
by minimizing the lattice energy for different lattice densities.  The minimum in $\Delta V_\mathrm{lat}$ for large lattice distances occurs for a ratio $L_\mathrm{x}/L_\mathrm{z}\approx {1/3}$. This result is consistent with the  increase of the inter-DQW attractive potential with respect to the repulsive one(cf.~Fig.~\ref{FigU}) in this parameter range. As a result, the total energy reduces if the first neighbors approach the central IX (i.e., for negative values of $\epsilon_\mathrm{r}$).
The change in sign of of $\epsilon_\mathrm{r}$ for decreasing densities is related to the fact  interaction energy $U_t$   (cf.~Fig.~\ref{FigU}) also changes sign and becomes positive when  $L_x>L_z/\sqrt{3}$.

\begin{figure}[!thbp]
\includegraphics[width=0.95\columnwidth, keepaspectratio=true]{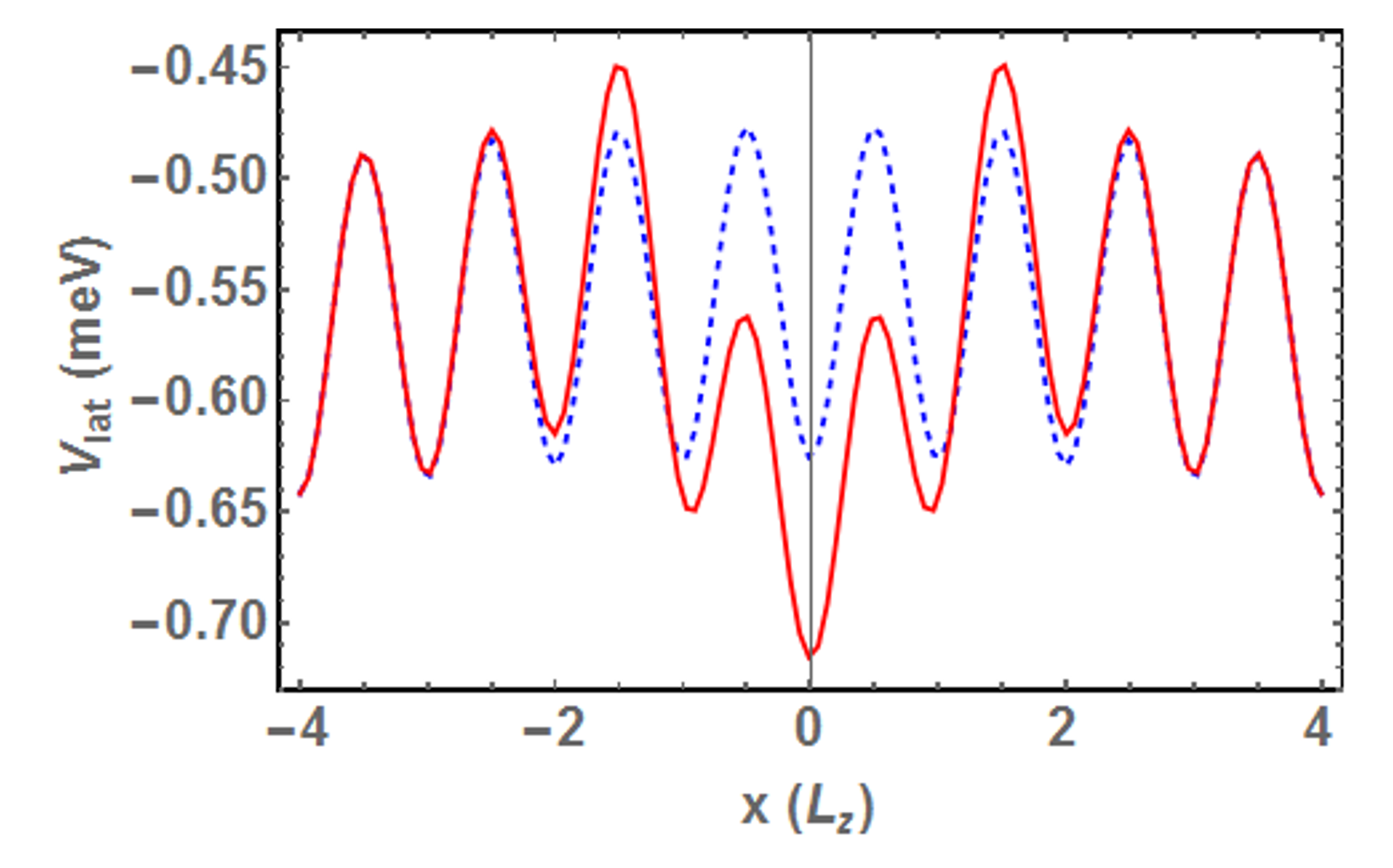}
\caption{Potential landscape across an IX \aPVS{di-}polaron. 
The solid line displays the IX potential calculated along the x-direction of the IX lattice \aPVS{di-}polaron displayed in the inset of Fig.~\ref{FigIXP2}. The calculations were carried out for the structure in the upper inset of Fig.~\ref{FigIXP1} with $L_\mathrm{x}=27.7$~nm, which corresponds to the minimum for $\Delta V_\mathrm{lat}$ in Fig.~\ref{FigIXP2}.
}
\label{FigIXP3}
\end{figure}

The  potential acting on the single IX in DQW$_2$ calculated at the minimum of Fig.~\ref{FigIXP2} are displayed in Fig.~\ref{FigIXP3}. The lattice deformation (of a few \%) required for \aPVS{di-}polaron formation is quite small. Also, the energy gain only reaches a few tens of \ensuremath{\mu}eV and is thus smaller than the typical spectral width of IX lines (on the order of a meV). The \aPVS{di-}polaron binding energy may be increased by increasing the IX dipole length or decreasing the spacing between the DQWs. The changes, however, are not expected to be drastic. As an example, we estimated that a reduction of the DQW spacer from 10~nm to 8~nm enhances the \aPVS{di-}polaron binding energy by approx. 30\%.

\subsubsection{Continuous \aPVS{Di-}polaron}
\label{Continuous Polaron}

The interaction between a single IX in DQW$_2$ and the excitonic liquid in DQW$_1$ \added[id=PVS]{in the inset of Fig.~\ref{fig:Theory}} can be also described in terms of a continuous, Fr\"ohlich-type \aPVS{di-}polaron, as given by the following Hamiltonian~\cite{Devreese15}:
\begin{equation}
\label{H1}
\hat H = \frac{\hat{\vec{p}}^2}{2 M}  + \sum_\vec{k} \hbar\omega (k) \bed_\vec{k}  \be_\vec{k} +  \sum_\vec{k} U (k) (e^{-i \vec{k} \hat {\vec{r}}} \bed_\vec{k}  +e^{i \vec{k} \hat {\vec{r}}} \be_\vec{k} ),
\end{equation}
with $\sum_\vec{k} = (2 \pi)^{-2} \int d^2k$. The first term of Eq.~\eqref{H1} describes the single excitonic ``impurity'' in DQW$_2$ with momentum $\hat{\vec{p}}$ and mass $M$; the second term gives the kinetic energy of the bosonic excitations (e.g.\ phonons) in the DQW$_1$ exciton liquid, as parametrized by the dispersion relation $\omega (k)$; and the last term represents the impurity-boson interactions. The momentum-space interaction parameter, $U(k)$, can be represented as:

\begin{equation}
\label{ukfkvk}
U (k) = f(k) V(k).
\end{equation}  

\noindent Here,
\begin{equation}
 V(k) = \int d^2 r U_{dd}(r) e^{- i \mathbf{k r}} = - \mu^2 \pi k~e^{-kL}
\end{equation}  

\noindent is the Fourier transform of the two-body interaction potential
\replaced[id=PVS]
{ $U_{dd}(r)$ given by Eq.~(\ref{EqUdd}) and  $\mu$ is defined in Eq.~(\ref{Eqmu}). }
{, $V(r) = \frac{\mu_1 \mu_2}{(L_z^2 + r^2)^{3/2}}\left(1 - \frac{3L_z^2}{L_z^2 + r^2} \right)$, 
and  $\mu_{1,2} = p_{1,2}/\sqrt{4 \pi \varepsilon \varepsilon_0} = e d_{1,2}/\sqrt{4 \pi \varepsilon \varepsilon_0}$.
}
The function $f(k) = \left[n_\mathrm{IX} \varepsilon(k)/(\hbar \omega(k))\right]^{1/2}$  depends on the density $n_\mathrm{IX}$, single-particle energy $\varepsilon(k) = \hbar k^2/(2m)$ as well as on the  correlation state of the IX gas expressed in terms of its dispersion relation $\hbar\omega(k)$. $m = m_e + m_{hh}$ is the exciton mass.

If the excitonic impurity in DQW$_2$ is static, it corresponds to an  ``infinite-mass \aPVS{di-}polaron'', $M=\infty$ (we set the excitonic impurity at $\vec{r} = 0$, for convenience). In this case, the Hamiltonian in Eq.~\eqref{H1} can be \added[id=PVS]{exactly} diagonalized using a coherent-state transformation

\begin{equation}
\hat S = \exp \left[- \sum_\vec{k} \frac{U(k)}{\hbar\omega(k)} \left(\bed_\vec{k} - \be_\vec{k}  \right) \right],
\end{equation} 
which gives the following ground-state energy shift:
\begin{equation}
\label{E0}
\Delta E_\mathrm{IX}  =  - \sum_\vec{k} \frac{U(k)^2}{\hbar\omega(k)} .
\end{equation} 

\noindent The corresponding ground eigenstate  is given by $\ket{\psi}= \hat S \ket{0}$, where $\ket{0}$ is the boson vacuum. Equation~\eqref{E0} gives a deformation energy of DQW$_1$ due to the interaction with a single exciton in DQW$_2$. In the case of a Hamiltonian with a linear coupling, such as Eq.~\eqref{H1}, the deformation energy is always negative, thus yielding a stable \aPVS{di-}polaron phase.

The ground state energy in Eq.~(\ref{E0}) depends on the correlation state of the IX fluid. Let us consider two limiting cases depending on the fluid density in  DQW$_1$: that of a low-density non-interacting exciton gas and of a high-density interacting exciton liquid. For a non-interacting exciton gas, the dispersion relation is that of free particles, $\hbar\omega(k) \equiv \varepsilon(k) = \hbar^2 k^2/(2m)$, thus resulting in a deformation energy given by:

\begin{equation}
\label{E02}
\Delta E_\mathrm{IX}  =  - n_\mathrm{IX}   \mu^4 \frac{\pi m}{\hbar^2L_z^2}.
\end{equation} 

For an interacting exciton gas or liquid, the dispersion relation can be stated as $\omega(k) \sim c (n_\mathrm{IX}) k$, where $c (n_\mathrm{IX})$ is the speed of sound in the IX fluid, which depends on the density $n_\mathrm{IX}$. One then obtains
\begin{equation}
\label{E03}
\Delta E_\mathrm{IX}  =  - n_\mathrm{IX} \mu^4 \frac{3\pi}{8 L_z^4 m c^2(n_\mathrm{IX})}.
\end{equation}

\noindent In the regime of Bogoliubov excitations, $c(n_\mathrm{IX}) \sim \sqrt{n_\mathrm{IX}}$, so that  $\Delta E_\mathrm{IX}$ becomes density-independent. The numerical computations of Ref.~\onlinecite{LozovikSSC07} revealed that the exciton liquid is strongly-interacting yielding  $c(n_\mathrm{IX}) \sim n_\mathrm{IX}^{0.7}$ (see Fig. 3(b) of Ref.~\onlinecite{LozovikSSC07}). One then obtains an energy shift $\Delta E_\mathrm{IX} \sim -n_\mathrm{IX}^{-0.4}$ with magnitude decreasing with increasing density.

\begin{figure}[!tbhp]
\includegraphics[width=.95\columnwidth, keepaspectratio=true]{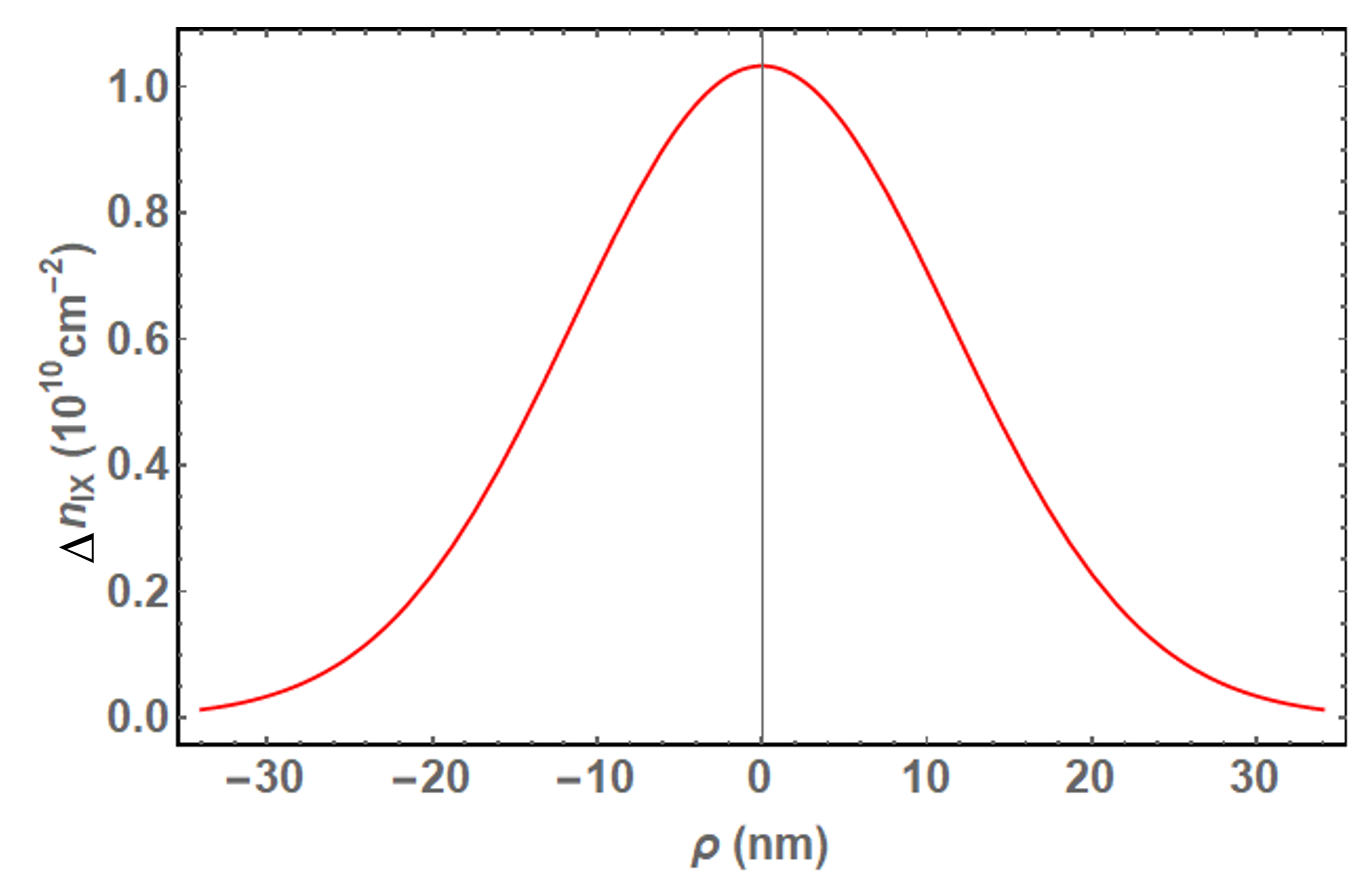}
\caption{\aPVS{Di-}polaron density profile $\Delta n_\mathrm{IX}$ determined from Eq.~(\ref{DeltaN}) for a fluid density $n_\mathrm{IX}=10^{10}$~cm$^{-2}$ and $L_z=34$~nm.
}
\label{FigPolProfile}
\end{figure}

The solid and dotted lines in Fig.~\ref{fig:Theory} compare the deformation energies predicted by  Eqs.~\eqref{E02} and \eqref{E03}, respectively, with the lattice \aPVS{di-}polaron model discussed in Sec.~\ref{IX Lattice Polaron}. One can see that the results for a correlated exciton liquid, Eq.~\eqref{E03}, are in good agreement with the lattice model at large densities $n_\mathrm{IX}$. The latter is  consistent with the fact that the lattice model presupposes correlations in displacements of the IX within the lattice. The uncorrelated model (solid line in Fig.~\ref{fig:Theory}) gives, in contrast, much larger energy red-shifts $|\Delta_\mathrm{IX}|$ with increasing densities. The interlayer binding energies predicted by the model for densities in the range $5\times10^{10}-10^{11}$ are indeed comparable to the experimentally measured ones.\cite{PVS315}  This agreement is, however, surprising since the high binding energies partially arise from the fact that the \aPVS{di-}polaron model assumes a free-particle  IX dispersion and, thus, neglects repulsive intra-DQW interactions, which are expected to become important with increasing densities.

In real space, the density deformation of DQW$_1$ is given by $\Delta n_\mathrm{IX} (\rho) = \bra{\psi} \bed_\vec{r}\be_\vec{r} \ket{\psi}$, where $\bed_\vec{r} = \int d^2 k/(2\pi)^2 \bed_\vec{k} e^{i \vec{k} \vec{\bf \rho}}$ and $\vec{\bf\rho}=\vec{r_\parallel}$. 
In the case of a correlated exciton fluid in DQW$_1$, the density deformation $\Delta n_\mathrm{IX} (\rho)$  at small values of $\bf\rho$ can be approximated by a Gaussian curve given by

\begin{equation}
\label{DeltaN}
\Delta n_\mathrm{IX} (\rho)  =  n_\mathrm{IX} \frac{ \mu^4}{2\hbar m c^3(n_\mathrm{IX})}   \frac{9\pi}{16 L_z^5 } e^{-\frac{\rho^2}{2 \sigma^2}},
\end{equation}

\noindent where $\sigma = 2 L_z/\sqrt{35}$. Note that the spatial extension of the \aPVS{di-}polaron cloud only  depends  on the inter-DQW spacing $L_z$ while the amplitude of the fluctuations reduce with increasing density $n_\mathrm{IX}$. For the experimental conditions of Ref.~\onlinecite{PVS315} with $L_z=34$~nm, the full width at half maximum of the deformation is 27~nm, as illustrated in Fig.~\ref{FigPolProfile}.
The fluid density almost doubles at the lateral position of the single IX. By integrating $\Delta n_\mathrm{IX}$ over the DQW plane one obtains a total density excess corresponding to approx. 0.1 particles for $n_\mathrm{IX}=10^{10}$~cm$^{-2}$.

\section{Conclusions}
\label{Conclusions}

We have theoretically investigated dipolar polarization effects in the interaction between IXs in stacked DQW structures. The dipolar inter-DQW coupling in these structures turns from repulsive for large lateral distances between the IXs to attractive for short separations.  The attraction enables the bonding of the IXs into an IX-molecule with a binding energy of a few hundreds of $\mu$eV,\cite{PVS284} as calculated in the point charge approximation for the IX dipoles.
We show that the dipolar inter-DQW interaction  deforms the wave function of the electron and hole constituents of the excitons with two main consequences: an increase in the IX binding energy (by approximately 20\%) as well as a dependence on the dipole moment on the amplitude of the applied field, thus yielding a finite IX polarizability.  

We have also analyzed multi-particle effects in the inter-DQW dipolar interaction by considering the coupling between a single IX in one of the DQWs to a (static) close-packed  lattice of IXs in the other DQW. For this configuration, the binding energy of the single IX reduces with increasing lattice density. This behavior is due to the dominating role of the intra-DQW dipolar repulsion, which prevents that more than one IX enters the attractive region of the inter-DQW coupling. We also found that the inter-DQW attraction can create a distortion of the lattice, thus forming an  IX \aPVS{di-}polaron and increasing the IX binding energy. \added[id=ML]{The lattice model was found to be in good agreement with the  theory for a continuous, Fr\"ohlich-like \aPVS{di-}polaron in the correlated regime.} \aPVS{This is not surprising due to the correlated nature of the dipolar interactions of the lattice models. In the low IX density, uncorrelated regime, the Fr\"ohlich-like \aPVS{di-}polaron model also accounts for the regime, where the measured interlayer binding energy increases with density. The simple approximation presented here neglects, however,   intra-DQW interactions and can, thus, only be regard as a rough approximation for the density regime in the transition from uncorrelated to correlated behavior. 
In fact, the interplay between the dipolar and the short-range exchange interactions at high particle densities has been recently shown to lead to the formation of mixed dark and bright  condensed Bose phases with interesting physics of interacting quantum liquids.\cite{High_Nature_12,Combescot_RoPiP80_66501_17,Stern_S343_55_14,Cohen_NL16_3726_16,Mazuz-Harpaz_PNAS116_18328_19}}
We expect nevertheless that the results presented here will stimulate further experimental and theoretical studies of dipolar inter-DQW interactions in IX systems consisting for DQW stacks.

{\bf Acknowledgments:} We thank W. Kaganer for discussions and for comment on the manuscript. We acknowledge the financial support from the German-Israeli Foundation (GIF), grant agreement I-1277-303.10/2014.
 \added[id=ML]{M.L.~acknowledges support by the Austrian Science Fund (FWF), under project No.~P29902-N27, and by the European Research Council (ERC) Starting Grant No.~801770 (ANGULON). }
 \added[id=AG]{A.G.~acknowledges support by the European Union’s Horizon 2020 research and innovation program under the Marie Skłodowska-Curie grant agreement No 754411.}
 \aPVS{P.V.S acknowledges financial support from the Deutsche Forschungsgemeinschaft (DFG) under Project No. SA 598/12-1.}


\def\litdir{}
\IfFileExists{x:/sawoptik_databases/jabref/literature.bib}
{   \def\litdir{x:/sawoptik_databases/jabref} }
{	\def\litdir{c:/myfiles/jabref} }

\IfFileExists{c:/users/santos/onedrive/latexIamHere.dat}
{   \def\onedrivedir{c:/users/santos/onedrive} }
{	\def\onedrivedir{y:/storage/onedrive} }

\IfFileExists{literature.bib}
{   \def\litdir{.} }
{	\def\litdir{.} }


%

%

\end{document}